\documentclass[preprint]{aastex631}

\usepackage{amsmath,amstext}
\usepackage{mathtools}
\usepackage{bm}
\usepackage{color}

\accepted{November 9, 2021}

\shorttitle{Spontaneous formation of magnetically driven outflows from accretion flows}
\shortauthors{Takasao et al.}
\graphicspath{{./}{figures/}}
\begin{document}

\title{Spontaneous Formation of Outflows Powered by Rotating Magnetized Accretion Flows \\
in a Galactic Center}

\author[0000-0003-3882-3945]{Shinsuke Takasao}
\affiliation{Department of Earth and Space Science, Graduate School of Science, Osaka University, Toyonaka, Osaka 560-0043, Japan}

\author{Yuri Shuto}
\affiliation{Kagoshima University, Graduate School of Science and Engineering, Kagoshima 890-0065, Japan}

\author[0000-0002-8779-8486]{Keiichi Wada}
\affiliation{Kagoshima University, Graduate School of Science and Engineering, Kagoshima 890-0065, Japan}
\affiliation{Ehime University, Research Center for Space and Cosmic Evolution, Matsuyama 790-8577, Japan}
\affiliation{Hokkaido University, Faculty of Science, Sapporo 060-0810, Japan}

\begin{abstract}
We investigate how magnetically driven outflows are powered by a rotating, weakly magnetized accretion flow onto a supermassive black hole using axisymmetric magnetohydrodynamic simulations. Our proposed model focuses on the accretion dynamics on an intermediate scale between the Schwarzschild radius and the galactic scale, which is $\sim$1-100 pc. We demonstrate that a rotating disk formed on a parsec-scale acquires poloidal magnetic fields via accretion and this produces an asymmetric bipolar outflow at some point. The formation of the outflow was found to follow the growth of strongly magnetized regions around disk surfaces (magnetic bubbles). The bipolar outflow grew continuously inside the expanding bubbles. We theoretically derived the growth condition of magnetic bubbles for our model that corresponds to a necessary condition for outflow growth. We found that the north--south asymmetric structure of the bipolar outflow originates from the complex motions excited by accreting flows around the outer edge of the disk. The bipolar outflow comprises multiple mini-outflows and downflows (failed outflows). The mini-outflows emanate from the magnetic concentrations (magnetic patches). The magnetic patches exhibit inward drifting motions, thereby making the outflows unsteady. We demonstrate that the inward drift can be modeled using a simple magnetic patch model that considers magnetic angular momentum extraction. This study could be helpful for understanding how asymmetric and non-steady outflows with complex substructures are produced around supermassive black holes without the help of strong radiation from accretion disks or entrainment by radio jets such as molecular outflows in radio-quiet active galactic nuclei, NGC~1377.
\end{abstract}



\section{Introduction}
Outflows of ionized and molecular gases are widely observed in various types of active galactic nuclei (AGNs) over a wide range of scales, from an accretion disk scale ($\sim 10^{-6}$ pc) to a galactic scale ($10^2$--$10^6$ pc) \citep[for example,][]{2008ARA&A..46..475H, fabian2012, 2014A&A...562A..21C}. Feedback owing to outflows is an essential process for the formation of supermassive black holes (SMBHs) and galaxies \citep{heckman2014,inayoshi2020}. Intense radiation from AGNs is a promising outflow driving mechanism \citep{king2015}, for example, the radiation pressure for the ionized gas \citep{proga2000,nomura_ohsuga2017, nomura2020} and the radiation pressure on the dusty gas \citep{ishibashi_fabian2015}. The radiation-driven outflows and failed winds may form obscured tori, which are essential for explaining type-1 and type-2 dichotomy in AGNs \citep{wada2012radiation, wada2015obscuring, willamson2020} (for review, see \citet{2015ARA&A..53..365N}).

In addition to the radiation from the nucleus, the magnetic field plays a key role in gas dynamics around supermassive black holes
\citep[for example,][]{kato2008bhad.book.....K,  ohsuga_mineshige2014SSRv..183..353O,hawley2015SSRv,Tchekhovskoy2015,yuan2014}. The magnetic field drives disk accretion via magneto-rotational instability \citep[MRI;][]{Balbus_Hawley_1991} and powers various types of outflows  \citep{1982MNRAS.199..883B,1985PASJ...37..515U,suzuki2009ApJ,bai2013ApJ,ohsuga2011,dihingia2021}. Many structures in the galactic center of our Galaxy are attributed to magnetic fields: a helical structure \citep{morris2006}, molecular loops \citep{fukui2006,machida2009}, noncircular motion of the gas \citep{suzuki2015,kakiuchi2018}, radio arcs \citep{sofue2005,morris2015}, and spurs \citep{sofue1977,kataoka2021}.

Many lines of observational evidence suggests the presence of approximately poloidal magnetic fields with a strength of $10~ {\rm \mu G}$--$1~{\rm m G}$ in our Galactic center \citep[for review, see]{ferriere2009interstellar}. The parsec-scale magnetic field structure in the circumnuclear disk (CND) is inferred from dust polarization data \citep{hsieh2018}. 
The event horizon telescope polarimetry imaging has recently estimated the strength of the magnetic field near the event horizon of the SMBH ($r \sim 7\times 10^{-4}$~pc) in M87 to be 1--30~G \citep{The_Event_Horizon_Telescope_Collaboration2021-id}. This observation supports the idea that the striking radio jet in M87 is magnetically accelerated, at least in part. These observations suggest the presence of magnetic fields at various scales in galactic centers. However, the origin of the magnetic field is unknown. The magnetic fields in galactic centers may be brought from galactic disks or generated inside accretion disks via the disk dynamo \citep{hawley2015SSRv}.

In addition to highly ionized outflows, cold molecular outflows have been observed in galactic centers.
\cite{aalto2016, aalto2017} observed a bipolar molecular outflow using Atacama Large Millimeter/submillimeter Array (ALMA) CO (3-2) in a Compton-thick AGN, NGC 1377. The bipolar outflow comprises narrow jets and wider (possibly conical) winds. The CO emissions were non-uniform along the molecular jets, suggesting internal substructures. The projected lengths of both the Northern and Southern parts were $\sim 150$~pc. 
In contrast to the radio jets observed in radio-loud AGNs, this outflow comprises a cold, dense molecular gas. Neither the radiation-driven outflow scenario nor the entrainment scenario wherein unseen radio jets entrain cold gas to form the molecular outflows are applicable in this low luminosity (the AGN luminosity is approximately 2\% of the Eddington luminosity), extremely radio-quiet galaxy.

Molecular outflows are commonly observed in star-forming regions \citep{arce2007,frank2014}. The rotation of outflows, a strong indication of magnetic acceleration, has also been observed \citep[for example,][]{bjerkeli2016resolved,hirota2017NatAs}. 
Theoretical and numerical studies on outflows at different stages of star formation are generally consistent with observations \citep{tomisaka1998ApJ...502L.163T, anderson2003ApJ,machida2008ApJ,joos2012,tomida2013ApJ,tsukamoto2018ApJ}. 
Considering the similarity to the molecular outflow in NGC 1377, \citet{aalto2020} suggested that the peculiar outflow in this galaxy could be powered by magnetic fields associated with gas accretion onto the CND. A theoretical investigation of this possibility is required. In particular, how such a complex outflow is formed by magnetic fields remains elusive.

Multi-dimensional MHD simulations are powerful tools for studying gas dynamics in galactic central regions.
For example, the fine turbulent structure of a parsec-scale magnetized molecular torus has recently been investigated using three-dimensional (3D) magnetohydrodynamic (MHD) simulations \citep{kudoh_wada2020ApJ...904....9K}.
However, most previous MHD simulations of CNDs assumed kinematically equilibrium initial conditions with pre-described magnetic fields. The configuration of the initial magnetic field is often assumed to be purely poloidal or toroidal \citep{machida2013ApJ...764...81M, chan_krolik2017ApJ...843...58C,kudoh_wada2020ApJ...904....9K,dorodnitsyn2017ApJ...842...43D}. If the magnetic fields in galactic centers are brought from the galactic scale through mass accretion, the magnetic field strength and configuration in the disk should be determined by the dynamical accretion process. This indicates that structures on an intermediate scale between the galactic and accretion disk scales impose a boundary condition for the growing disk outflow system.
It is theoretically suggested that the mass accretion from the
galactic scale toward the galactic center forms a CND at $r \sim$ 1-100~pc \citep[for example,][and references therein]{kawakatu_wada2008ApJ...681...73K}.
This is also confirmed via recent high-resolution observations using ALMA in nearby AGNs \citep{combes2019A&A...623A..79C, garciaburillo2021galaxy}.
Especially, the central region of the type-2 Seyfert galaxy NGC 1068 
has been extensively studied using various molecular lines. Observations demonstrate that the molecular lines of the CND in NGC 1068 show counter-rotation and that the CND is connected to the surrounding structures \citep{garcia-burillo2019A&A...632A..61G, imanishi2020ApJ...902...99I}, thereby implying that gas accretion to the galactic central region is key physics for determining the structure and kinematics of the CND.

In this paper, we study spontaneous formation of magnetically driven outflows from a growing CND in a galactic center. In other words, we investigate how magnetically driven outflows are formed without the help of feedback from the AGN or starburst. We numerically solved the accretion of a magnetized gas from a 100-pc scale. Motivated by the observations of the 100-pc scale molecular outflows in NGC~1377 \citep{aalto2020}, our numerical model covers the spatial scale ranging from 0.1--300~pc. With this approach, we aim to solve the following fundamental questions for our model. 1) How does the central pc-scale disk acquire magnetic fields during its growth? 2) How and when are magnetically driven outflows powered from the growing disk? 3) What physical processes can produce substructures in outflows?

The remainder of this paper is organized as follows. Section~\ref{sec:method} describes the model setting, numerical methods, and boundary conditions. 
The numerical results of the disk formation process and the formation of magnetic outflows are presented in Section~\ref{sec:results}. We will show the spontaneous formation of an asymmetric outflow with complex substructures.
In Section~\ref{sec:discussion}, we summarize and discuss the formation process of such an outflow. A theoretical explanation for the growth condition of outflows is also provided. Section~\ref{sec:conclusions} summarizes our conclusions regarding the three key questions raised above.

\section{Numerical Setup}\label{sec:method}
\subsection{Model Setting}\label{subsec:model}
Our model is axisymmetric (2D) and is constructed in spherical coordinates $(r,\theta)$. As described below, our model is greatly simplified, and a direct comparison of our simulation with observations is beyond the scope of this study. In this study, we focused on basic MHD processes relevant to magnetically driven outflows. To study the formation of outflows from a growing CND, we performed a long-term ($\sim 2.5\times 10^4$ orbital periods at the inner boundary) simulation. We faced many numerical challenges during the long-term evolution. We describe our methods to overcome the difficulties.

In our model, the central black hole is located at the center of the spherical coordinates and it behaves as a point mass with a mass of $M_{\rm BH}$. The gravitational potential in our model is produced only by this central black hole. The gravitational potential of the bulge is ignored. We simulated the formation of a rotating disk via accretion of the surrounding magnetized gas. We resolved the Bondi radius, inside which the gravitational energy dominates the thermal energy of the gas. 
The Bondi radius $r_{\rm B}$ is defined as 
\begin{align}
    r_{\rm B} = \frac{GM_{\rm BH}}{c_{s,\infty}^2}
\end{align}
where $G$ is the gravitational constant, and $c_{s,\infty}$ is the sound speed corresponding to $T_{\infty}$.
To avoid artificial assumptions for the initial and boundary conditions for the disk gas and magnetic field distributions as much as possible, we simulated the disk formation by modeling the accretion from such an intermediate scale. The accretion of an unmagnetized gas was studied in detail using the approaches of \citet{inayoshi2019MNRAS} and \citet{Sugimura_Hosokawa_Yajima_Inayoshi_Omukai_2018}, where the physical outer boundary of the accretion disk is set at the Bondi radius. We extend such a hydrodynamic model to an MHD model to understand the origin of magnetically driven outflows.

The normalization unit of our model is summarized in Table~\ref{tab:units}. The mass of the central black hole is assumed to be $9.0\times 10^6~{\rm M_\odot}$, which is equivalent to the estimated SMBH mass of NGC~1377 \citep{aalto2020}. The radius of the inner boundary $r_{\rm in}$ was assumed to be 0.1~pc, while the outer boundary was located at $3,000\,r_{\rm in}=300$~pc. The Bondi radius $r_{\rm B}$ was set to $1,000\, r_{\rm in}=100$~pc. Therefore, there is a difference of three orders of magnitude between the inner boundary radius and the Bondi radius. The condition at the Bondi radius physically determines the boundary condition for the inner part. Our model focuses on the intermediate scale between the black hole Schwarzschild radius ($\sim 10^{-6}$~pc) and the galactic bulge scale ($\sim 1$~kpc). As a measure of velocity, we will use the local escape velocity $v_{\rm esc}=\sqrt{2GM_{\rm BH}/r}$.

\begin{deluxetable*}{ccc}
\tablenum{1}
\tablecaption{Normalization Units. $M_{\rm BH}=9.0\times10^{6}$ M$_{\odot}$ is adopted. \label{tab:units}}
\tablewidth{0pt}
\tablehead{
\colhead{Quantities} & \colhead{Units} & \colhead{Fiducial Values}
}
%\decimalcolnumbers
\startdata
        Length & $L_{0}=r_{\rm in}$ & 0.1~pc  \\
        Velocity & $v_{0}=v_{K0}$ & $6.22\times10^7$ cm s$^{-1}$ \\
        Time & $\displaystyle t_{0}=\frac{r_{\rm in}}{v_{K0}}$ & $5.0\times10^{9}$ s ($1.6\times 10^2$~yr)\\
        Keplerian orbital period & $\displaystyle t_{\rm K0}=2\pi t_0$ & $3.1\times 10^{10}$~s ($9.9\times 10^2$~yr) \\
        Density & $\rho_{0}$ & $1.00\times10^{-23}$ g cm$^{-3}$\\
        Magnetic field strength & $B_{0}=\sqrt{4\pi\rho_{0}}v_{K0}$ &$6.97\times10^{-4}$ G \\ 
        Mass accretion rate & $\dot{M_{0}}=\rho_{0}L_{0}^{3}/t_{0}$ & $9.41\times10^{-7}$ M$_{\odot}$yr$^{-1}$\\
        Temperature & $\displaystyle T_{0}=\frac{k_{B}}{\gamma m_{H}}\frac{GM_{BH}}{r_{\rm in}}$ & $4.46\times10^7$ K
\enddata
%\tablecomments{}
\end{deluxetable*}

\begin{figure}
    \centering
    \includegraphics[width=13cm]{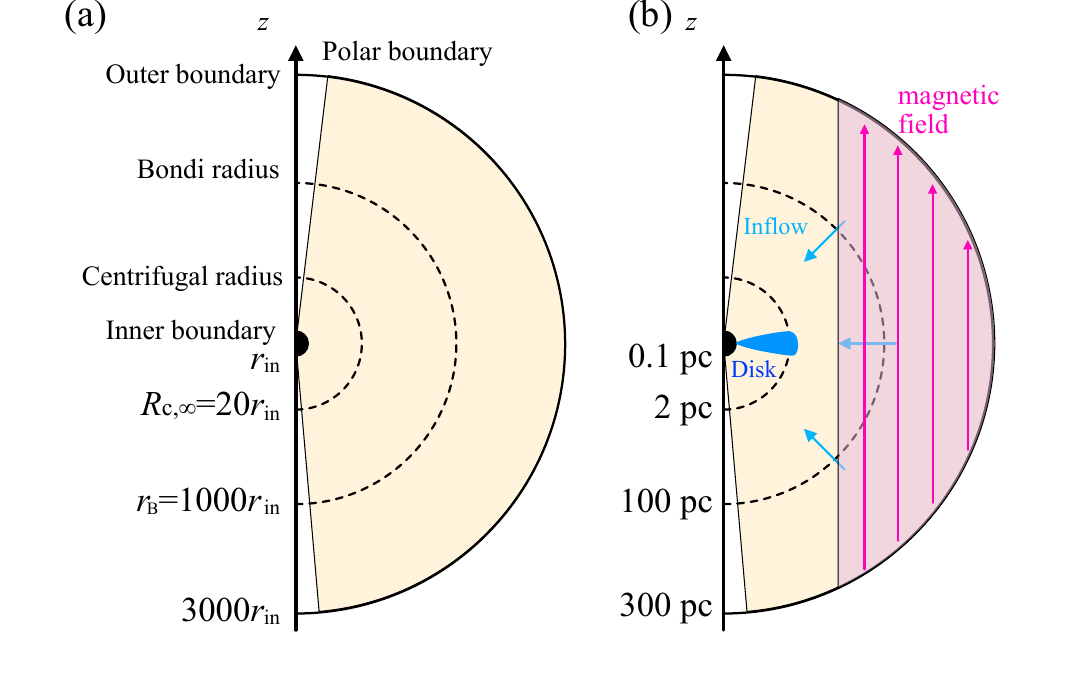}
    \caption{Schematic diagram of the model setting. (a) the initial condition, here the density is uniform and the rotational velocity is given by Equation~\ref{eq:ang}. The region inside the inner boundary is colored in black. (b) The gas and magnetic field structures at the time of magnetic field insertion.}
    \label{fig:model_setup}
\end{figure}

As shown in Figure~\ref{fig:model_setup}~(a), the black hole is initially surrounded by an unmagnetized gas having a uniform density $\rho_{\rm B}$ (density at the Bondi radius scale) and temperature $T_{\infty}$ and a spatially changing specific angular momentum $j$. $\rho_{\rm B}=\rho_0$ and $T_\infty=10^{-3}T_0$.
For simplicity, we adopt the equation of state for an ideal gas with a mean molecular weight of unity. Thus, $c_{\rm s,\infty}=\sqrt{\gamma k_{\rm B}T_\infty/m_{\rm H}}$, where $\gamma$ is the specific heat ratio, $m_{\rm H}$ is the hydrogen mass, and $k_{\rm B}$ is the Boltzmann constant.
To form a CND inside the numerical domain, we control the initial profile of the specific angular momentum. The initial profile is given as
\begin{align}
    j=\begin{dcases}
    \left(\frac{R}{r_{\rm B}}\right)^2 j_\infty & (R\ge r_{\rm B})\\
    j_\infty & (R < r_{\rm B}),\label{eq:ang}
    \end{dcases}
\end{align}
where $R=r\sin\theta$ is the cylindrical radius.
$j_\infty$ is defined by the centrifugal radius $R_{\rm c,\infty}$ as follows:
\begin{align}
    j_{\infty}^2 = GM_{\rm BH} R_{\rm c,\infty},
\end{align}
where $R_{\rm c,\infty}$ determines the typical size of the CND in the early phase of disk formation.
The centrifugal radius $R_{c,\infty}$ is $20r_{\rm in}=2$~pc. 

After the formation of the disk (defined as the time after 8,000 times the inner Keplerian orbital periods after the simulation starts), we impose a straight, uniform magnetic field outside the cylindrical radius of $55r_{\rm in}$ (larger than the disk size at this time), as shown in Figure~\ref{fig:model_setup}~(b). This is just to avoid numerical instabilities around the inner boundary. We characterize the magnetic field strength by the plasma $\beta$, which is the ratio of the gas pressure to the magnetic pressure. Using the gas pressure of the initial condition, we assume that the plasma $\beta$ of the magnetized region is $10^{6}$, which corresponds to $3\times10^{-2}$~$\mu$G. This field strength is too weak to affect the accretion dynamics. However, the large-scale field continuously accumulates in the disk as a result of accretion. Even if the initial magnetic fields are negligible for the accretion dynamics, the magnetic fields can become dynamically important as they accumulate. We studied this transitional phase.

\subsection{Basic Equations and Numerical Methods}
To study the accretion and outflow processes in our axisymmetric model, we solve the following normalized, resistive, and viscous MHD equations in a conservative form in spherical coordinates using Athena++ \citep{Stone_Tomida_White_Felker_2020}:
\begin{align}
        \frac{\partial \rho}{\partial t} + \nabla \cdot ( \rho \bm{v} )= 0 \\
        \frac{\partial \rho \bm{v}}{\partial t} + \nabla \cdot (\rho \bm{vv} - \bm{BB} + \mathrm{P^*} + \bm{\Pi}) = -\frac{GM_{\rm BH}\rho}{r^2}\bm{e}_r\\
        \frac{\partial e_{\rm tot}}{\partial t} + \nabla \cdot [(e_{\rm tot}+P^*)\bm{v} - \bm{B}(\bm{B \cdot v}) + \bm{\Pi \cdot v} + \eta \bm{J} \times \bm{B} ] = -n^2\Lambda \\
        \frac{\partial \bm{B}}{\partial t} - \nabla \times [(\bm{v} \times \bm{B}) - \eta \bm{J} ] = 0,
\end{align}
where $\rho$ is the density, $\bm{v}$ is the velocity vector, and $\bm{B}$ is the magnetic field vector. $\mathrm{P}^*$ is the diagonal tensor with components $p_{\rm tot} = p + {B}^2/2$, where $p$ is the gas pressure and $p_{\rm tot}$ is the total pressure. {$\bm{e}_{r}$ is the unit vector in the $r$-direction.} $\bm{J}=\nabla \times \bm{B}$ denotes the electric current density vector. $\eta$ is the resistivity, and it comprises the effective resistivity as a result of turbulence $\eta_{\rm eff}$ and an artificial resistivity $\eta_{\rm in}$, which is introduced to avoid numerical instabilities around the inner boundary, namely, $\eta(r,\theta)=\eta_{\rm eff}(r,\theta)+\eta_{\rm in}(r,\theta)$. The descriptions of $\eta_{\rm eff}$ and $\eta_{\rm in}$ will be provided later. $\bm{\Pi}$ denotes the viscous stress tensor.
\begin{align}
    \Pi_{ij} = \rho \nu \left( \frac{\partial \upsilon_{i}}{\partial x_{j}} + \frac{\partial \upsilon_{j}}{\partial x_{j}} - \frac{2}{3} \delta_{ij} \nabla \cdot \bm{v} \right).
\end{align}
In this study, viscosity is given by the effective viscosity in response to turbulence, $\nu_{\rm eff}$: $\nu(r,\theta)=\nu_{\rm eff}(r,\theta)$, defined at a later stage. 
$e_{\rm tot}$ is the total energy density, which is defined as 
\begin{align}
    e_{\rm tot} = e_{\rm int} + \frac{1}{2}\rho \upsilon^2 + \frac{B^2}{2},
\end{align}
{where $e_{\rm int}=p/(\gamma-1)$ is the internal energy density.} To close the equations, we use the equation of state for an ideal gas. Considering the effect of radiative cooling, the specific heat ratio of the gas $\gamma$ is assumed to be smaller than 5/3, and we adopted a value of 1.05. {$\Lambda$ is a radiative loss function and is only applied to the accretion-shocked region. The description will be given in Appendix}~\ref{sec:cooling}.

We adopt the second-order piecewise linear reconstruction method and the third-order strong stability preserving Runge--Kutta time integration method. The numerical flux is calculated using the Harten--Lax--van Leer Discontinuities (HLLD) approximate Riemann solver \citep{Miyoshi_Kusano_2005}.

The numerical domain covers the region of $(r_{\rm in}, \Delta \theta_{\rm B}) \le (r,\theta) \le (3,000 \, r_{\rm in},\pi-\Delta \theta_{\rm B})$, where $\Delta \theta_{\rm B}=\pi/180$. This region was resolved using $256\times 256$ meshes. The mesh spacing was uniform in the $\theta$ direction. To keep the ratio of the radial and latitudinal meshes nearly constant, we let the radial mesh spacing be proportional to the radius. The mesh size increase ratio is 3\%. Namely, if we consider i as the mesh index in the $r$-direction, the radial mesh size increases as $dr_{\rm i+1}/dr_{\rm i}=1.03$.
The radial and latitudinal mesh sizes at $r=R_{c,\infty}=20r_{\rm in}$ were $0.6r_{\rm in}$ and $7.3\times 10^{-3}r_{\rm in}$, respectively.

\subsection{Inner and Outer Boundary Conditions}
The inner boundary in the $r$-direction is an outgoing boundary where the flows toward the numerical domain are prohibited. To avoid numerical instabilities, we artificially enhance resistivity just around the inner boundary. The detailed treatment is described in Appendix~\ref{appendix:inner-boundary}.
At the outer boundary, the gas with the same density and temperature as those of the initial gas is continuously injected at a constant mass injection rate of $\dot{M}_{\rm B}$, where $\dot{M}_{\rm B}$ is the Bondi accretion rate. When we adopted the outflowing (zero-gradient) boundary condition for the hydrodynamic quantities, the gas outside the Bondi radius was depleted and the accretion rate at the Bondi radius became significantly smaller than the Bondi accretion rate. The zero-gradient boundary conditions are applied to the magnetic field. We assume the reflecting boundary conditions for the north and south poles. Therefore, the magnetic fields do not escape across the poles from the numerical domain.

\subsection{Effective Resistivity and Viscosity as a result of MRI Turbulence}\label{sec:effective_diff}
In weakly magnetized, non-self-gravitational disks, turbulence in response to MRI is believed to be the main driver of accretion \citep{Balbus_Hawley_1991}.
However, MRI turbulence is intrinsically a 3D process. In particular, magnetic reconnection of the azimuthal component of magnetic fields is essential to determine the saturation level of magnetic field amplification via MRI \citep[for example, ][]{Sano_Inutsuka_2001}.
As shown in Appendix \ref{appendix:MRI-model}, in axisymmetric, ideal MHD simulations, disk magnetic fields are subject to a continuous amplification and the disk plasma $\beta$ becomes close to or lower than unity as time proceeds, which is unlikely in reality. Therefore, to model MRI-turbulent disks using 2D axisymmetric models, we should consider the 3D effect.

Previous 3D simulations of MRI \citep{lesur2009} demonstrated that the effective resistivity is comparable to the effective viscosity in a wide range of parameters. Considering this, we phenomenologically modeled the effective viscosity $\nu_{\rm eff} $ and resistivity $\eta_{\rm eff}$ in response to the MRI turbulence:
\begin{align}
    \eta_{\rm eff}= \nu_{\rm eff}=\alpha c_{\rm s}H,\label{eq:eta_eff}
\end{align}
where $c_{\rm s}$ is the local isothermal sound speed, and $H=\sqrt{2}c_{\rm s}/\Omega_{\rm K}$ is the disk pressure scale height. $\Omega_{\rm K}$ is the Keplerian angular velocity. $\alpha$ is the disk viscosity parameter. Many previous simulations for different disk conditions indicate that $\alpha\sim \mathcal{O}(0.01)$-$\mathcal{O}(0.1)$ in the saturated state of MRI turbulence \citep[for example, ][]{hawley2013,Suzuki_Inutsuka_2014}. Hence, it is reasonable to expect that in the disk body, the effective viscosity $\nu_{\rm eff}$ with $\alpha\sim 0.01$ arises from the MRI turbulence.

To model both magnetic diffusion and viscosity arising from the MRI turbulence in our 2D axisymmetric model, we include effective resistivity and viscosity that only operate in the disk body. Our MHD model solves large-scale magnetic fields but small-scale, turbulent magnetic fields. As our model lacks small-scale magnetic field fluctuations producing turbulent viscosity, we explicitly include effective viscosity. Their values were determined according to the above estimates with $\alpha=0.01$. With this approach, we can avoid unphysical amplification of disk magnetic fields during long-term evolution with the help of effective resistivity. Simultaneously, despite the weakening of disk magnetic fields owing to the effective resistivity, the effective viscosity drives disk accretion with the accretion rate expected for MRI-turbulent disks. A more detailed description of our model is provided in Appendix~\ref{appendix:MRI-model}.

\section{Results}\label{sec:results}
\subsection{Disk Formation Process}
\begin{figure}
    \centering
    \includegraphics[width=16cm]{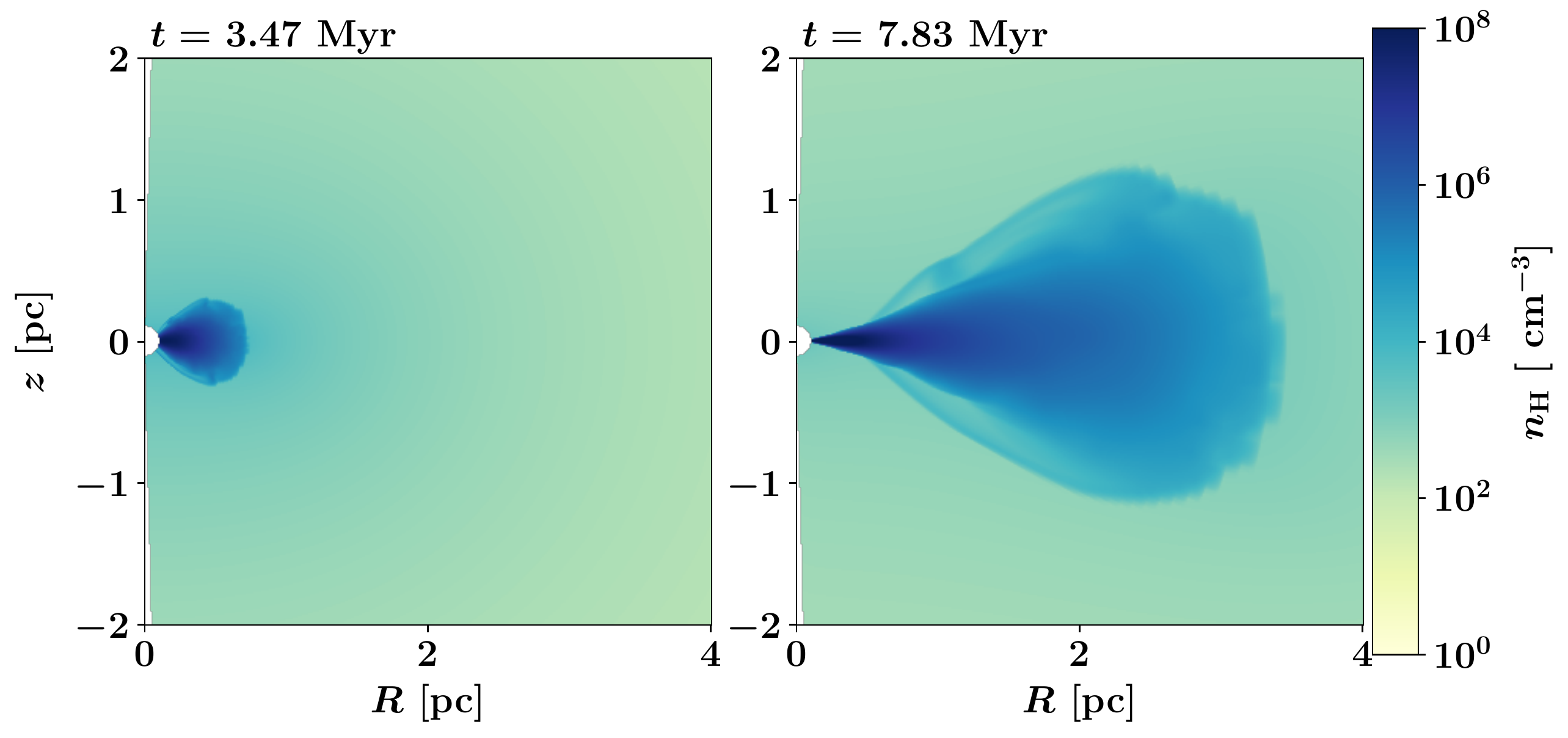}
   \caption{Formation of the gas disk as a result of mass accretion before the magnetic field is imposed. The density distributions at $t=3.47$~Myr (left) and $t=7.83$~Myr (right) are shown.}  %(Density.py)
  	\label{fig:Density2panel}
\end{figure}

Figure~\ref{fig:Density2panel} shows the formation of the unmagnetized disk. The disk size increases to $\sim 3$~pc up to 1.2~Myr, which is comparable to the centrifugal radius (2~pc). The disk surface is exposed to the supersonic accretion flows. The density discontinuity around the disk is the accretion shock. 

\begin{figure}[t]
\begin{center}
\includegraphics[width=10cm]{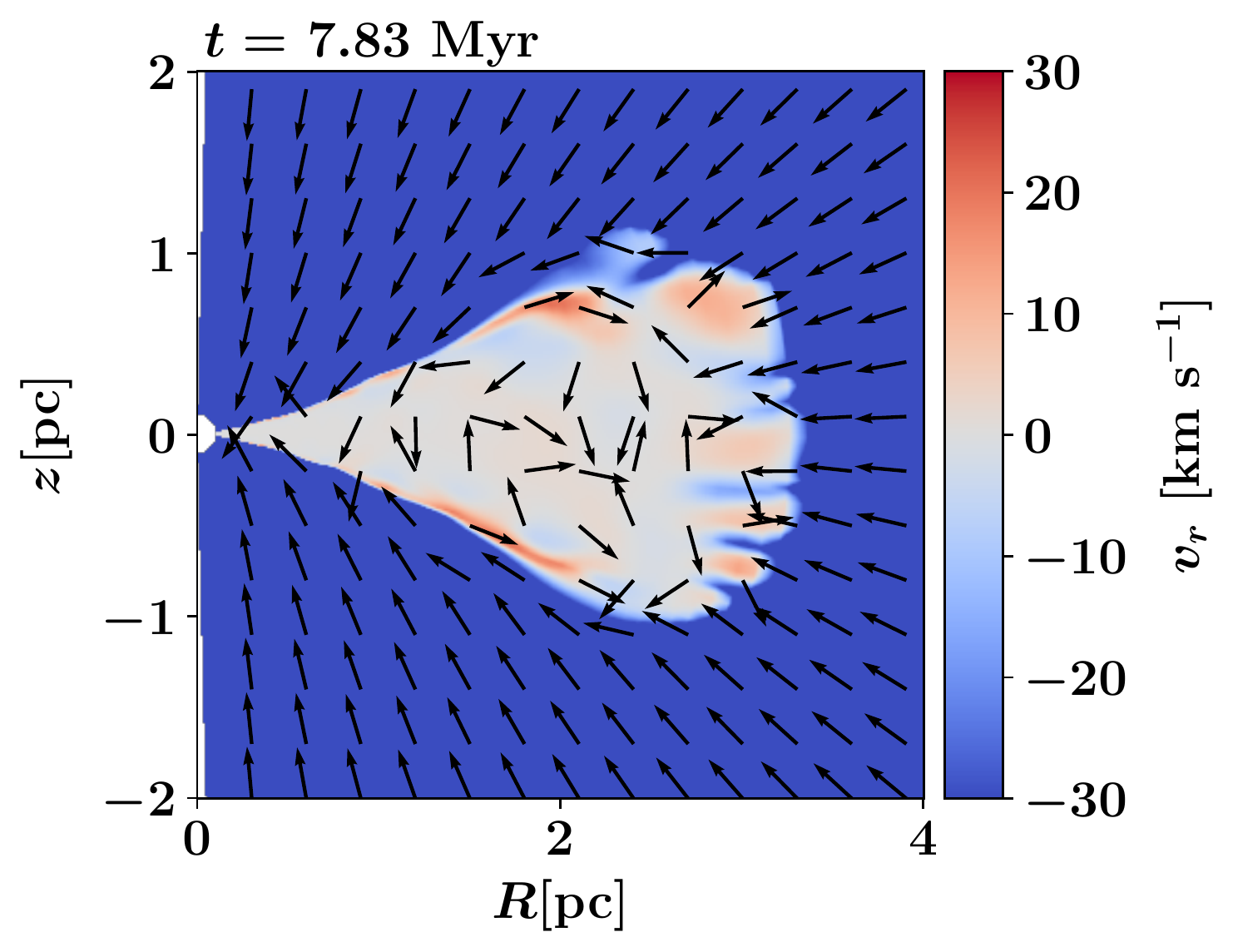} %(VrVk2Dplot.py)
\caption{Radial velocity $v_r$ and velocity vector in the $R$-$z$ plane at $t=7.83$~Myr. 
Regions with negative $v_r$ correspond to the accreting flows.} \label{fig:vr_conv}
\end{center}
\end{figure}

We overview the velocity structure around the disk surface, as it plays an important role in magnetic field transport. Figure~\ref{fig:vr_conv} shows the ratio of the radial velocity to the local Keplerian velocity, $v_r/v_{\rm K}$. We note that magnetic fields have not yet been added. The arrows denote the direction of the velocity vectors (the arrow size does not indicate the speed). It is shown that nearly free-falling gas hits the disk. After the flows pass through the accretion shocks, they change their directions to be toward the center. As a result, the nearly free-falling accretion flows form layers behind the accretion shocks at the disk surfaces in both hemispheres. Hereafter, we call this flow as \textit{the disk surface accretion}. A similar structure was observed in \citet{takasao2021}, where the accretion around a proto-gas giant was simulated.

Looking at the outer edge, we obtain plumes that penetrate the disk. The penetrating plumes enhance the gas pressure and drive outgoing flows in the disk, which produces a vortical flow pattern. The formation of the penetrating plumes is related to the relative size of the disk to $R_{\rm c,\infty}$. The disk size in the early phase is comparable to $R_{\rm c,\infty}$, but it becomes considerably larger than $R_{\rm c,\infty}$ owing to the angular momentum transport inside the disk. As a result, when the accreting gas falls onto the outer edge, the gas is not rotationally supported and moves radially inward, resulting in the formation of plume-like flows. Owing to this disturbance, the outer edge of the disk becomes highly asymmetric about the equatorial plane.

\subsection{Transport and Accumulation of Magnetic Fields}
When the unmagnetized disk was developed ($t= 8000t_{\rm K0}\approx 7.9$~Myr), a uniform magnetic field was imposed, as described in Section~\ref{subsec:model} (see Figure~\ref{fig:model_setup}). We provide an overview of the transport and accumulation of magnetic fields in the disk.

\begin{figure}[t]
\centering
\includegraphics[width=18cm]{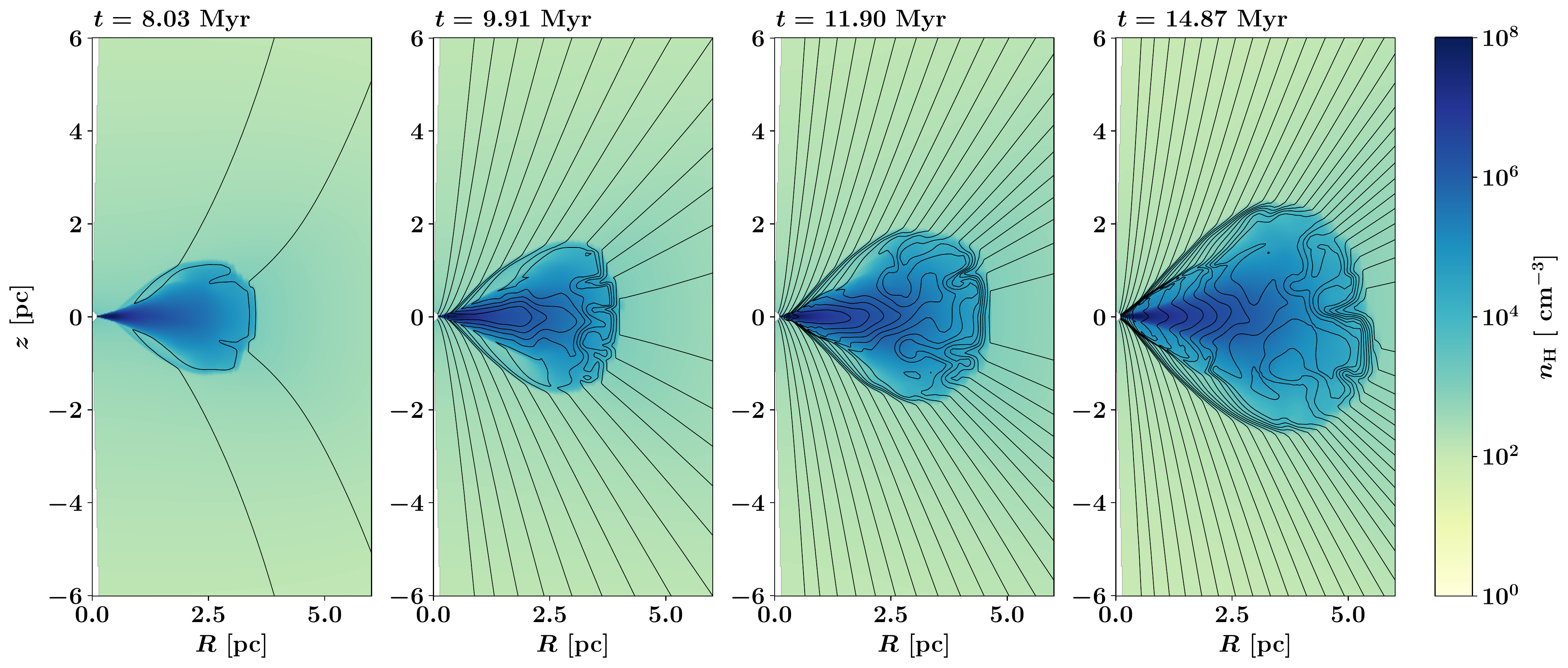}
\caption{Evolution of the magnetic field and the gas disk after the magnetic field is imposed at $t= 8$ Myr. The black lines denote magnetic field lines. The color indicates the number density of hydrogen.}\label{fig:density_bline}
\end{figure}

Figure~\ref{fig:density_bline} displays the evolution of magnetic field lines around the disk. The color shows the density. One will observe that magnetic fields are significantly dragged toward the center around the disk surfaces. The disk surface accretions just behind the accretion shocks efficiently drag the magnetic fields. The speed of the disk surface accretions is comparable to the local Keplerian velocity (Figure~\ref{fig:vr_conv}), but the accretion velocity at the midplane is much smaller. The importance of the disk surface accretion for the magnetic flux transport has been reported in several studies \citep{Matsumoto_1996,Beckwith_2009,Takasao_2018,Takasao_2019}. The radial transport of the poloidal fields around the midplane is mediated by magnetic diffusion.
In Figure~\ref{fig:density_bline}, we can also obtain the invasion of finger-like magnetic fields, which is caused by the plume-like flows at the outer edge of the disk (Figure~\ref{fig:vr_conv}).

\begin{figure}[t]
\centering
\includegraphics[width=10cm]{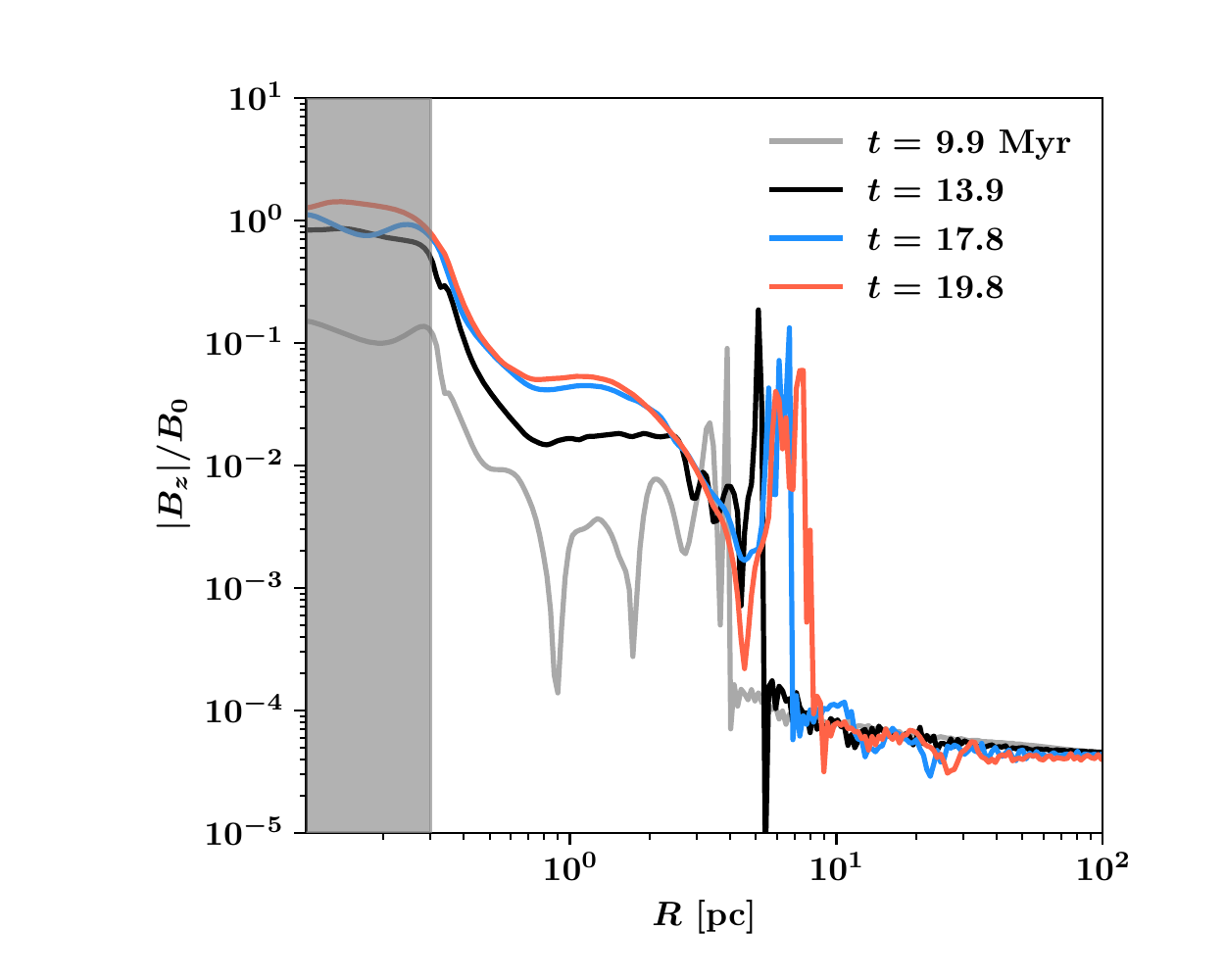}
\caption{Evolution of the radial distribution of $|B_z|$ at $z=0$ after the magnetic field is imposed. In the hatched region ($R < 3$ pc), magnetic diffusion was implemented.} \label{fig:BzRslice}
\end{figure}

Figure~\ref{fig:BzRslice} shows the temporal evolution of the radial distribution of $|B_z|$ on the equatorial plane. After the insertion of the magnetic fields, the disk poloidal field increases from the inner part. Later, $|B_z|$ is, nearly uniformly, enhanced in the disk body, which is caused by the transport via the gradual magnetic diffusion. We note that magnetically driven outflows are not observed until $t\approx 20$~Myr. As we will show in Section 3.3 and discuss in Sections 4.1 and 4.2, it is necessary for launching the outflows that the disk acquires a sufficient amount of poloidal magnetic flux.

\subsection{Magnetically driven Outflows}
We describe how magnetic outflows are launched from the developing magnetized disk.

\subsubsection{Development of Magnetic Bubbles and Outflows}
\begin{figure}
\centering
\includegraphics[width=18cm]{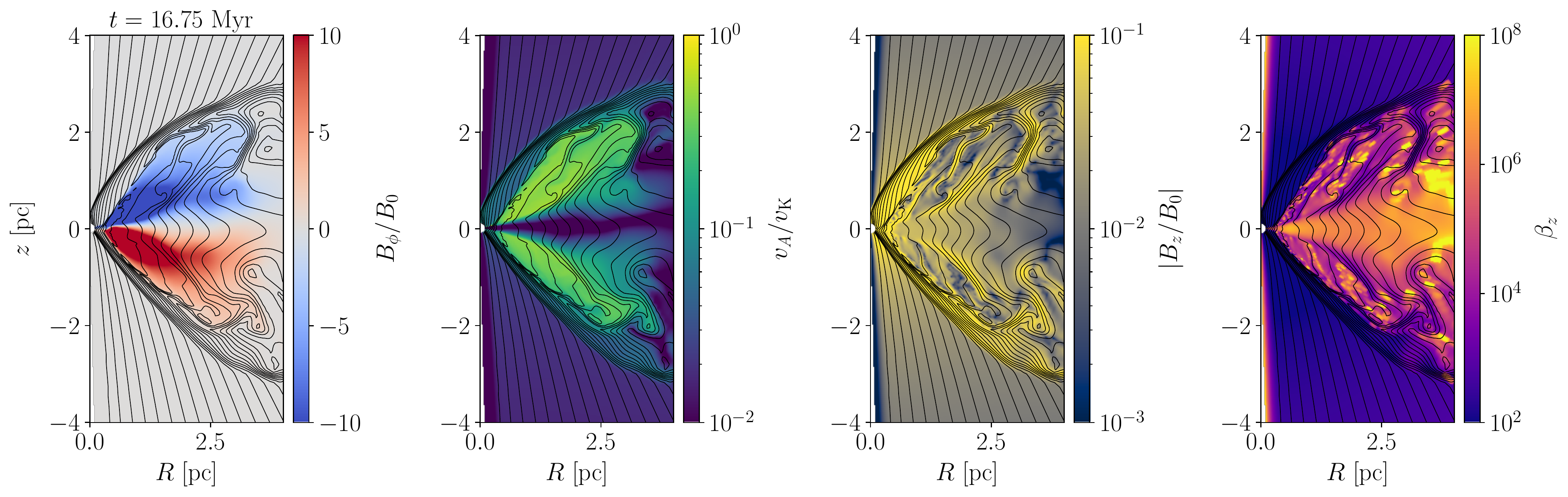}
\caption{Left to right, distributions of $B_\phi/B_0$, the ratio of the Alfv\'en speed and the local Kepler rotational velocity $(v_{\rm K}=\sqrt{GM_{\rm BH}/r})$, $v_{\it A}/v_K$, $|B_z/B_0|$, the plasma $\beta_z \equiv p/p_{{\rm mag}, z}$, where $p_{{\rm mag}, z}$ is the magnetic pressure based on $B_z$. The black lines indicate the magnetic field lines.}\label{fig:mag_va_beta}
\end{figure}

Along with the build-up of the disk poloidal fields, the toroidal component $B_\phi$ is amplified around the disk surfaces. The simulation shows that the strong $B_\phi$ regions gradually expand in both hemispheres before outflow growth. Both upflows and downflows are intermittently excited inside them in space and time. We call such structures ``magnetic bubbles'' and investigate their development.
Figure~\ref{fig:mag_va_beta} shows the early phase of the development of the magnetic bubbles. $B_\phi$ is amplified around the disk surfaces. 
The second left panel shows the ratio of the Alfv\'en speed $v_{\it A}$ to the local Keplerian velocity $v_{\rm K}$, indicating the local energy density ratio. This panel shows that the magnetic energy density inside the magnetic bubbles is comparable to the gravitational energy density.
The right panel shows the plasma $\beta$ based on the $B_{z}$ component only, where the inhomogeneity in the poloidal fields is highlighted. The field line structure is complicated owing to the mixture of upflows and downflows.
Magnetic bubbles are different from a so-called magnetic tower jet \citep{lynden-bell1996MNRAS,y-kato2004ApJ}. The outflow velocity structure is coherent in the magnetic tower jet, whereas outflows and downflows coexist in the magnetic bubbles. Therefore, the magnetic bubbles and outflows do not coincide in general, especially in the early phase of outflow growth. The magnetic structure in the magnetic bubbles is more complex than that of the magnetic tower jet as the outflows and downflows stir magnetic fields.

$B_\phi$ in the magnetic bubbles is generated as follows:
The disk surface accretion drags poloidal fields inward, producing $B_{R}$, where $B_{R}$ is the cylindrical radius component of the magnetic field. $B_R$ is then converted into $B_\phi$ via disk shear. This amplification mechanism is commonly observed in previous MHD simulations of accretion disks \citep[for example, ][]{Zhu_Stone_2018}. We emphasize that this mechanism operates efficiently, especially for the developing disk of this kind, as the disk surface accretion with a nearly free-fall velocity persists and continuously produces $B_R$ before the development of magnetic bubbles. 
As we will see later, the fast surface accretion disappears after the magnetic bubbles expand largely above the disk. 
Subsequently, the generation of $B_\phi$ from $B_z$ by the disk twisting motion becomes important (Section~\ref{sec:growth_condition}).

We denote the important role of magnetic bubbles in the formation of outflows. After the formation of magnetic bubbles, accretion shocks were formed at the outer edges of the bubbles. This indicates that the magnetic bubbles protect the disk surfaces against accretion flows. The accretion flows try to prevent the formation of outflows by pushing them down, but magnetic bubbles create spaces for the outflows to develop. Therefore, the formation of magnetic bubbles is essential for outflow growth.
Compared to the $B_z$ and $B_\phi$ maps in Figure~\ref{fig:mag_va_beta}, $B_\phi$ is dominant in the magnetic bubbles, which indicates that outflow is driven mainly by the magnetic pressure gradient force \citep{Shibata_1985, Kudoh_1998}.

\begin{figure}
    \centering
\includegraphics[width=16cm]{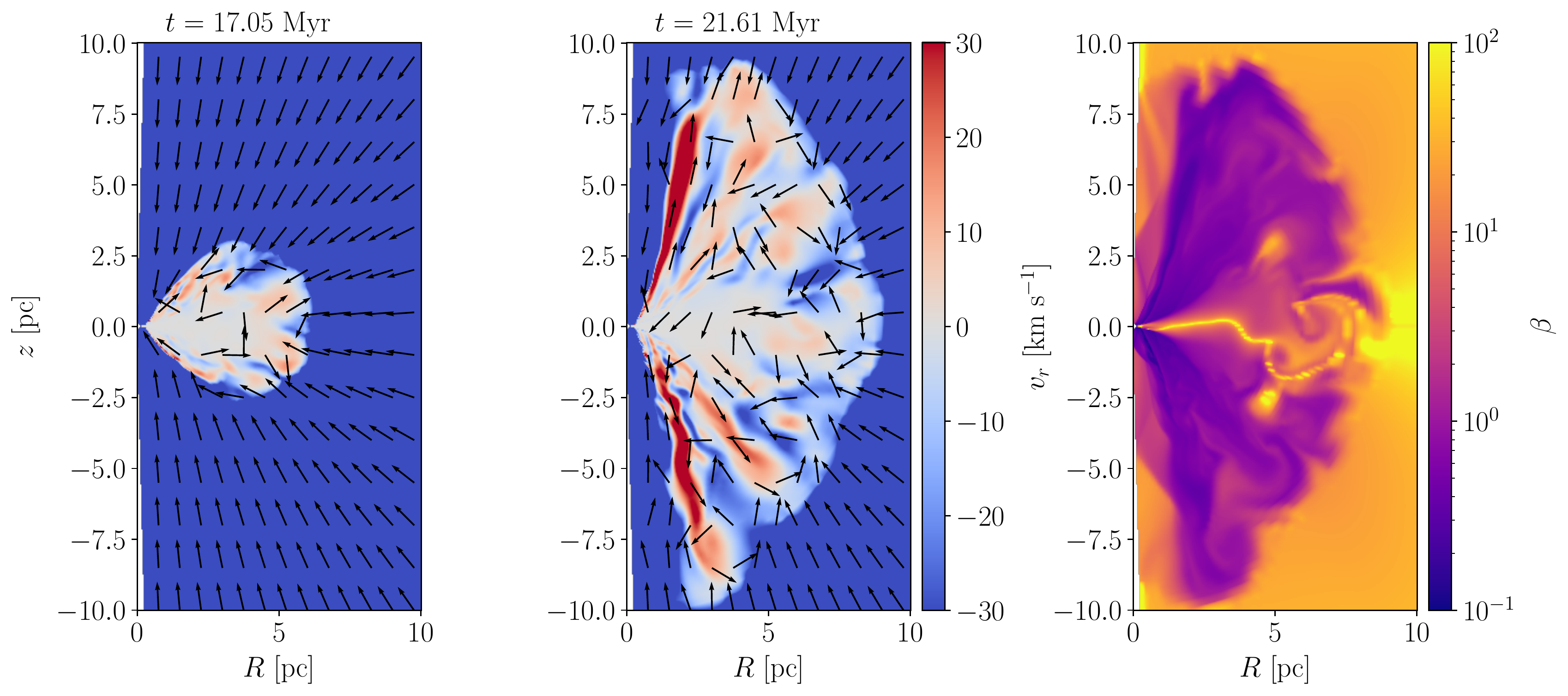}
    \caption{Growth of the magnetic bubbles and the mini-outflows inside them. The left and middle panels show the radial velocity $v_r$ at $t$=17.05~Myr (left) and $t$=21.61~Myr, respectively. The right panel displays the plasma $\beta$ at $t$=21.61~Myr.}
    \label{fig:onset_bubble}
\end{figure}

The magnetic bubbles show rapid growth around $t\approx 20$~Myr. Figure~\ref{fig:onset_bubble} displays the radial velocity $v_r$ and plasma $\beta$ structures around the disk at approximately the time of rapid expansion of the magnetic bubbles. At $t=17.05$~Myr (left panel), the magnetic bubbles are confined by the accreting flows around the disk surface. However, at a later time, the bubbles first grew into a half-spherical shape and extended further in the $z$ direction (middle panel). Upflows and downflows coexist in the magnetic bubbles because a portion of the outflow gas falls back onto the disk surfaces. The plasma $\beta$ in the majority of the bubbles is smaller than 0.3 (right panel), indicating strong magnetization. We will describe how the outflow grows as the magnetic bubbles expand in the next Section.

\subsubsection{Asymmetric Outflow with Patchy Substructures}
\begin{figure}
\centering
\includegraphics[width=16cm]{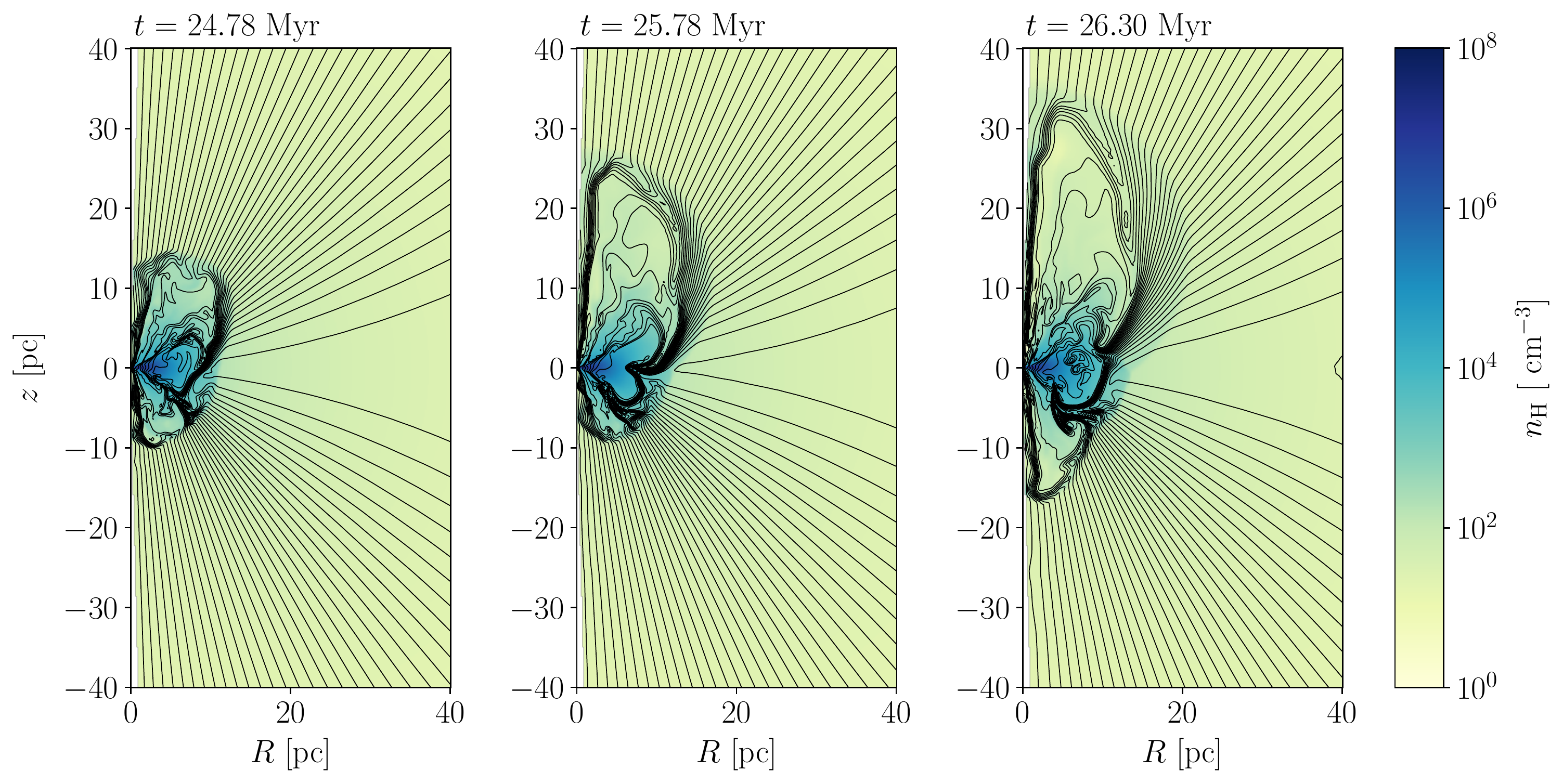}
\includegraphics[width=16cm]{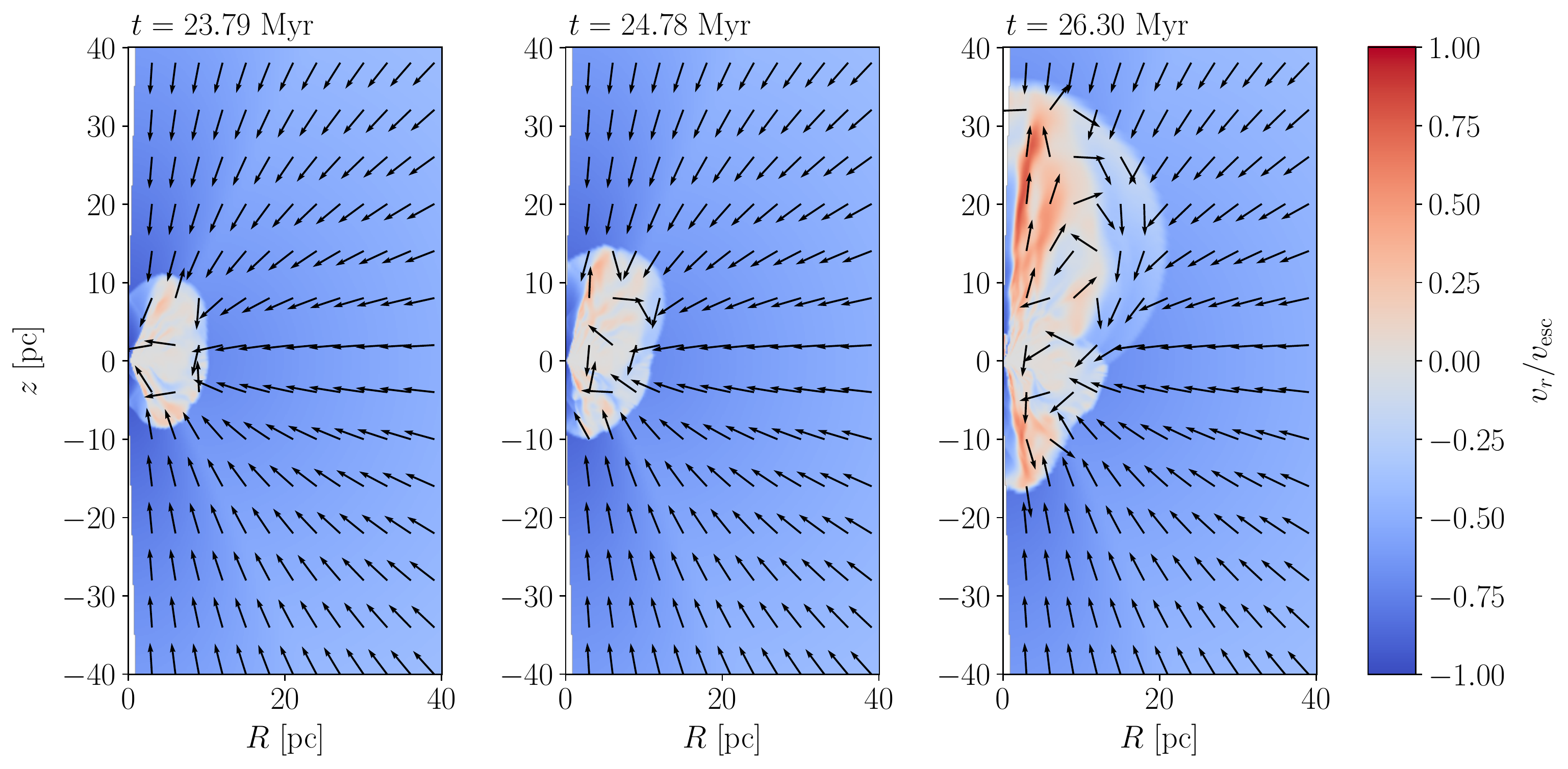}
 \caption{Formation of the bipolar outflow. The top panels show the density evolution with magnetic field lines, while the bottom panels display the evolution of the radial velocity normalized by the local escape velocity, $v_{\rm esc}=\sqrt{2GM_{\rm BH}/r}$. $v_{\rm esc}(30 {\rm pc})\approx 52$ km s$^{-1}$.}
    \label{fig:formation_outflow}
\end{figure}

Figure~\ref{fig:formation_outflow} displays the expansion of magnetic bubbles and the development of outflows. The top panels show the density with magnetic field lines, while the bottom panels indicate the ratio of the radial velocity to the local escape velocity, $v_r/v_{\rm esc}$. The arrows in the bottom panels show the direction of the poloidal velocity. An asymmetric bipolar outflow is launched inside the magnetic bubbles. 
The bipolar outflow is driven mainly by the magnetic pressure of the toroidal field $B_\phi$. At $t=26.3$~Myr, the outflow speed is close to the local escape velocity in some parts (at $r=30$ pc, $v_{\rm esc}\approx 52$ km s$^{-1}$), suggesting that a portion of the outflow will eventually escape from the central black hole gravity. We had to stop the calculation at $t\approx 27$~Myr owing to numerical instabilities around the inner boundary, which prevented us from investigating the final fate of the outflows.

\begin{figure}
    \centering
    \includegraphics[width=18cm]{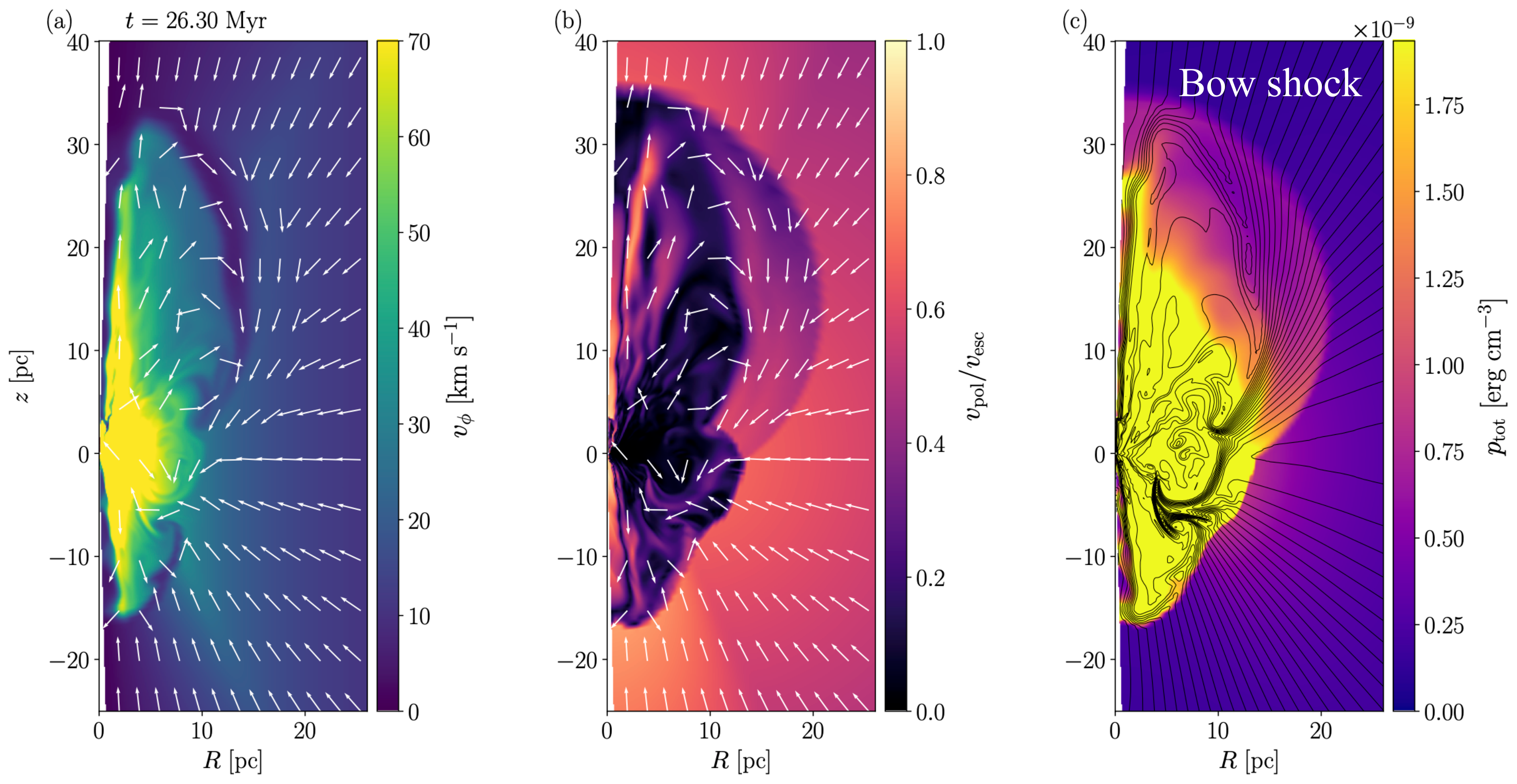}
    \caption{Structure of the bipolar outflow around at the end of the simulation. Panels ~(a) and (b) show $v_\phi$ and $v_{\rm pol}/v_{\rm esc}$, respectively. The poloidal velocity is defined as $v_{\rm pol} \equiv \sqrt{v_r^2 + v_\theta^2}$, and the local escape velocity is $v_{\rm esc} = \sqrt{2G M_{BH}/r}$. The arrows denote the direction of the velocity vectors (the arrow size does not indicate the speed). Panel~(c) indicates the total pressure distribution. The lines denote the poloidal magnetic field lines.}
      \label{fig:rotation_jet}
\end{figure}

Figure~\ref{fig:rotation_jet} shows the magnetic bubble structure at $t=26.3$~Myr. Panel~(a) shows the rotational velocity $v_\phi$, while Panel~(b) exhibits the poloidal velocity normalized by the local escape velocity, $v_{\rm pol}/v_{\rm esc}$. The panels show that the outflow structure is highly patchy, but the rotational profile is relatively smooth inside the magnetic bubbles. 
Panel~(a) indicates significant  reduction  in  the  rotational  velocity  at  the  outer  edge  of  the  bubbles. We show in Appendix~\ref{appendix:torque} that the reduction is caused by the magnetic force at the outer edge of magnetic bubbles. The shock structure is shown in Panel~(c), where the total pressure distribution is shown. A bow shock was formed around the expanding bubbles. The bow shock compresses gas and magnetic fields and changes the direction of accreting flows.

We describe the development of asymmetric bipolar outflow. Recall that the disk is disturbed by the plume-like flows at the outer edge (Figures~\ref{fig:vr_conv} and \ref{fig:density_bline}). This disturbance makes the disk accretion asymmetric about the midplane, resulting in the asymmetric growth of the magnetic bubbles.
In this simulation, the magnetic bubble in the Northern hemisphere grew faster than that in the Southern hemisphere.
As a result, the rapidly growing magnetic bubble prevented the growth of bubbles in the other hemisphere. The bottom panels of Figure~\ref{fig:formation_outflow} show the velocity structure around the magnetic bubbles at different times. The panels indicate that the accreting flows are reflected around the outer edge of the Northern hemisphere and are directed toward the Southern hemisphere. 

\begin{figure}
    \centering
    \includegraphics[width=8cm]{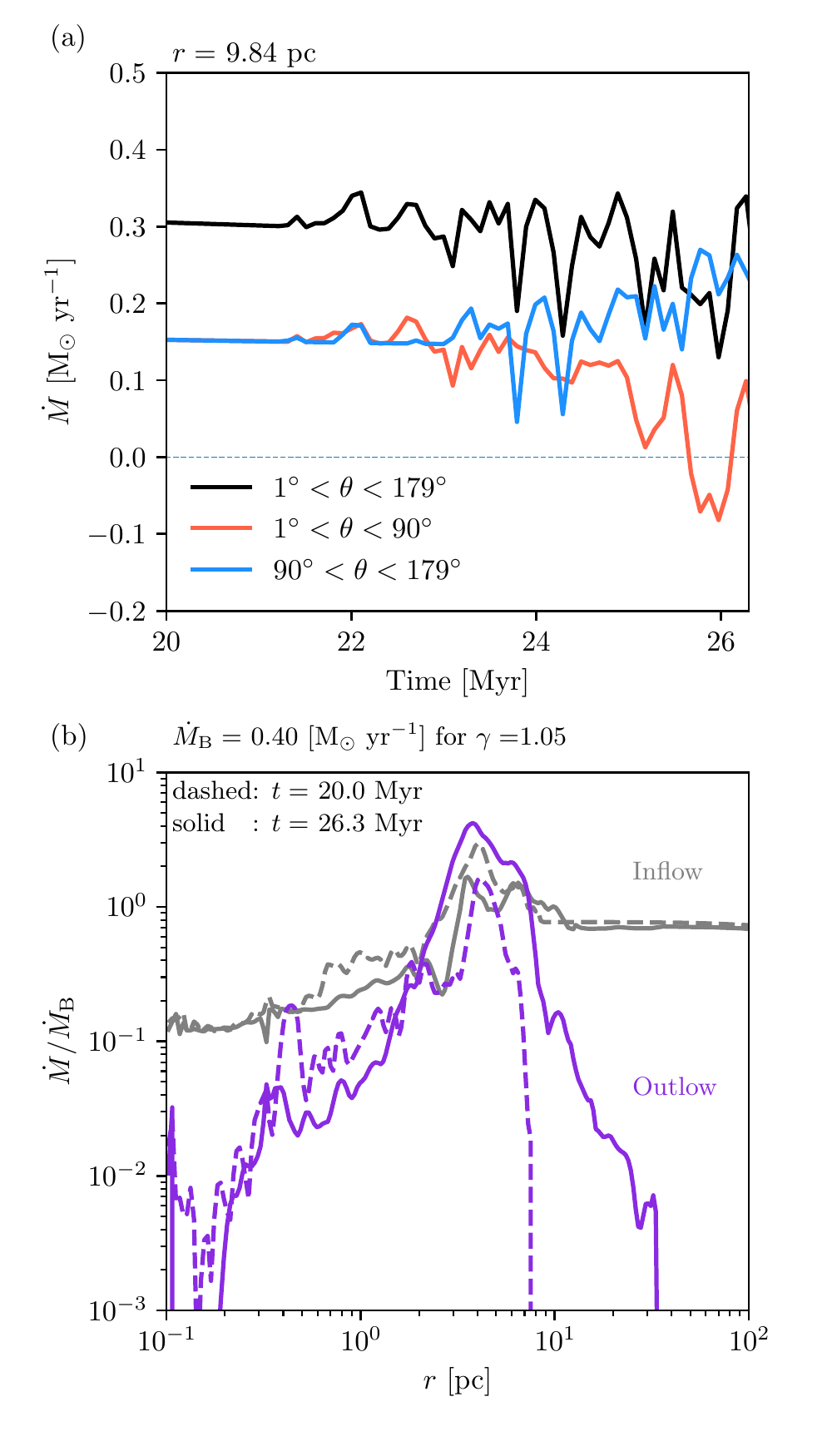}
    \caption{(a) Evolution of the mass inflow rate measured at 98$r_{\rm in}=0.98~{\rm pc}$. The black line shows the total mass inflow rate, while the red and blue lines denote the mass inflow rates in the Northern and Southern hemispheres, respectively. (b) Mass inflow and outflow rates as functions of radius. The values were normalized by the Bondi accretion rate $\dot{M}_{\rm B}=0.40~{\rm M}_\odot~{\rm yr}^{-1}$ (for $\gamma=1.05$). The gray and purple lines indicate the inflow and outflow rates, respectively. The dashed and solid lines show the data for $t=20.0$ and 26.3~Myr, respectively.}
    \label{fig:mass_inflow_rate}
\end{figure}

The asymmetric accretion structure resulted in an imbalance in the accretion rate in the Northern and Southern hemispheres.
The imbalance is displayed in panel~(a) of Figure~\ref{fig:mass_inflow_rate}. The red and blue lines show the mass accretion rates in the Northern and Southern hemispheres, respectively, while the black line denotes their sum. The accretion rate was measured at $r=98r_0=9.8$~pc. The accretion rates in the Northern and Southern hemispheres were obtained by integrating the mass flux within the ranges of $1^\circ < \theta < 90^\circ$ and $90^\circ < \theta < 179^\circ$, respectively. The accretion rates in both hemispheres are almost identical until $t\sim 3.7$~Myr, but they behave differently subsequently. The accretion rate in the Southern part of the disk increased, while it decreased in the Northern hemisphere. As the total accretion rate is nearly constant on average, the plot shows that a part of the accreting gas in the Northern hemisphere flows toward the Southern hemisphere. As a result, the ram pressure of the accretion flows increased in the Southern hemisphere, which prevented the development of the magnetic bubble.

Figure~\ref{fig:mass_inflow_rate}~(b) shows the mass inflow and outflow rates as a function of radius in spherical coordinates. 
The values were normalized by the Bondi accretion rate $\dot{M}_{\rm B}=0.40~{\rm M}_\odot~{\rm yr}^{-1}$ for $\gamma=1.05$. The inflow rate outside the disk ($\sim 10$~pc) was almost the same as the Bondi accretion rate and remained unchanged over time. The inflow and outflow rates are comparable around the outer edge of the disk ($r\sim 10$~pc) owing to counter-flows excited by plumes (Figure~\ref{fig:vr_conv}). Although they exceed the Bondi rate, the outgoing flows in the counter-flow region are confined by the gravitational potential owing to the SMBH, and the high outflow rate does not indicate the production of strong outflows. Inside the disk, the accretion rate is regulated by the effective viscosity, and it decreases by a factor of 10 ($\sim 0.1 \dot{M}_{\rm B}$). The contribution of the outflow is seen outside $r\sim 10$~pc at $t=$26.3~Myr. The purple solid line indicates that the outflow rate is $\sim 10^{-2}\dot{M}_{\rm B}$. A short summary of the mass flows is as follows: the mass is supplied to the disk at a rate of $\dot{M}_{\rm B}$. A fraction of the mass accreted to the center through the disk at a rate of $\sim 0.1 \dot{M}_{\rm B}$. The magnetically driven outflow carries the mass from the disk at $\sim 10^{-2}\dot{M}_{\rm B}$. Therefore, the ratio of the outflow rate to the accretion rate in the disk was approximately 0.1. The rest of the supplied mass accumulates at $\sim 0.9\dot{M}_{\rm B}$ and is used for disk growth.

\subsubsection{Drifting Motions of Mini-outflows}
The origin of the patchy substructures of outflows (mini-outflows) can be understood as follows. We observe that the $B_z$ distribution around the disk surfaces is highly non-uniform and time variable (see Figure~\ref{fig:mag_va_beta}). As $B_z$ is the source of both $B_R$ and $B_\phi$, the non-uniform $B_z$ distribution leads to a spatially varying Poynting flux. The mini-outflows are a natural result of such a non-uniform Poynting flux.

\begin{figure}
    \centering
    \includegraphics[width=18cm]{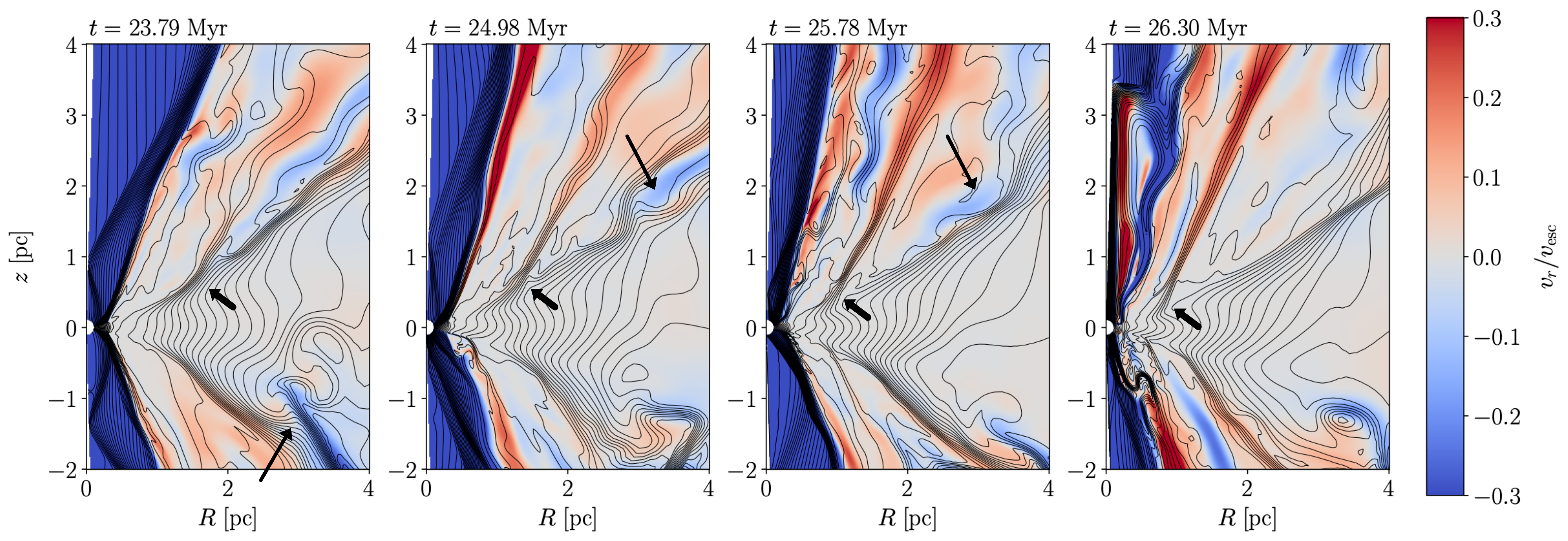}
    \caption{Example of magnetic patches drifting toward the center. The same magnetic patch at four different timings is indicated by the thick arrows. The color shows the radial velocity normalized by the local escape velocity. The black lines indicate poloidal magnetic field lines. The three thin arrows indicate regions where the fallback flows produce a non-uniform poloidal field distribution.}
      \label{fig:example_magnetic_patch}
\end{figure}

We obtain the radial motions of the bases of the mini-outflows. Hereafter, we refer to the bases of mini-outflows as ``magnetic patches'' according to \citet{Spruit_Uzdensky_2005} (hereafter, SU05). Figure~\ref{fig:example_magnetic_patch} shows an example of the drifting magnetic patches. The color shows $v_r/v_{\rm esc}$, and the black lines show poloidal magnetic field lines. The concentration of the poloidal fields around the disk surface indicated by the thick arrow is the drifting magnetic patch. The magnetic patch produces a faster outflow than its surroundings, while it drifts inward. We observe that the drifting motions contribute to the time variability of the outflows.

\begin{figure}
    \centering
    \includegraphics[width=18cm]{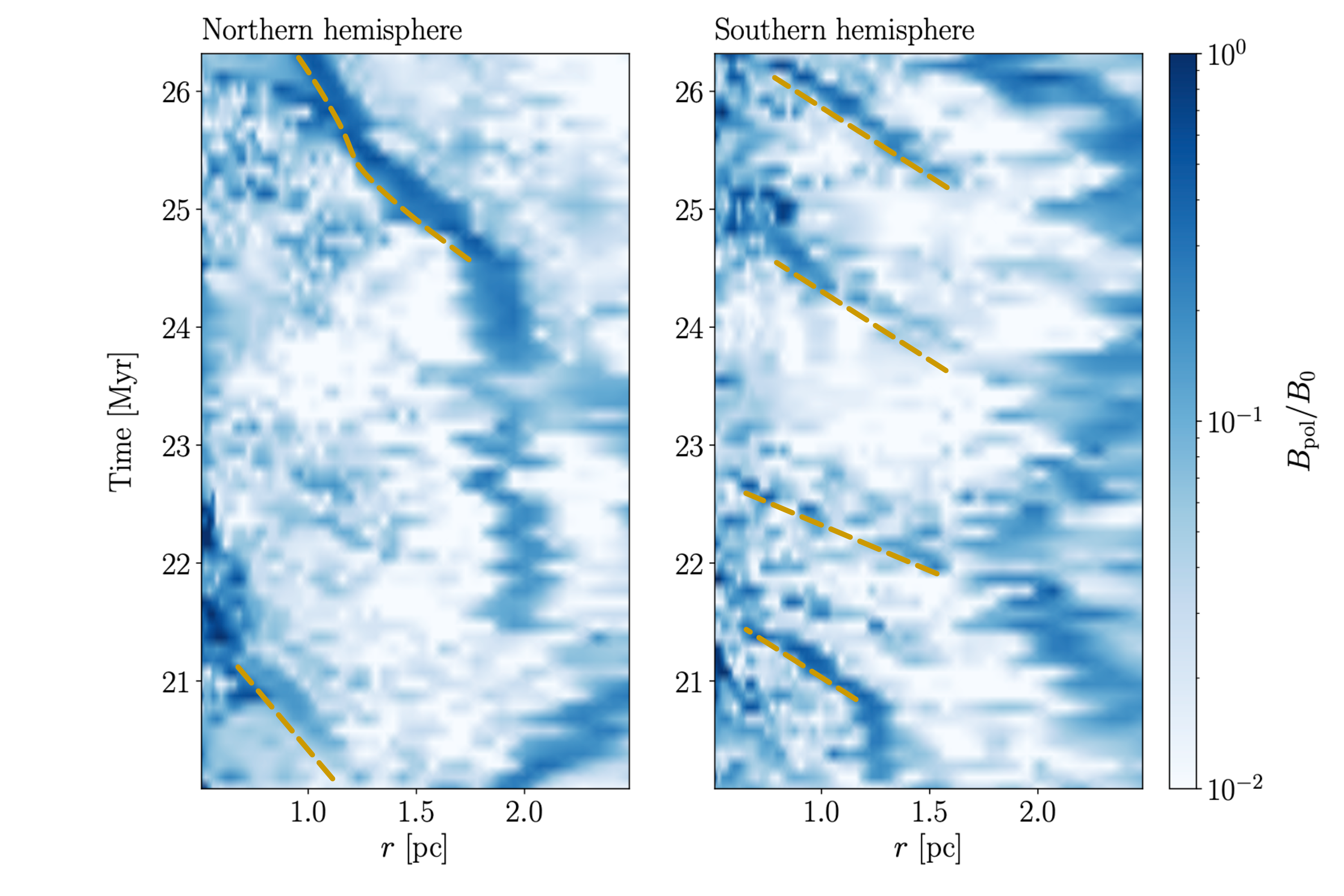}
    \caption{Drift motions of magnetic patches. Time--radius diagrams of the poloidal magnetic field strength at the disk surfaces for the Northern hemisphere (left) and for the Southern hemisphere (right) are shown. Color indicates the strength of the poloidal magnetic field strength. The distributions for the Northern and Southern hemispheres were measured along the spherical radius direction at angles of $60^\circ$ and $120^\circ$ from the north pole, respectively. Magnetic patches are defined as concentrations of poloidal fields (their drifting motions are indicated by dashed lines). (see also the right panel of Figure \ref{fig:schematic-diagram}).}
      \label{fig:magnetic_patches_NS}
\end{figure}

Figure~\ref{fig:magnetic_patches_NS} highlights the drifting motions of several magnetic patches. The figure shows the radius--time diagram of the radial distribution of the poloidal field strength, $B_{\rm pol}$. 
Color indicates the strength of the poloidal magnetic field strength. The distributions for the Northern and Southern hemispheres were measured along the spherical radius direction at angles of $60^\circ$ and $120^\circ$ from the north pole, respectively. Magnetic patches are seen as concentrations of poloidal fields, and their drifting motions are indicated by dashed lines.
The typical speed is approximately $1\times 10^{-3}L_0/t_0 \approx 0.6~{\rm pc~Myr^{-1}}$. Therefore, if the disk size is of the order of the parsec-scale, the drifting motions will introduce time variability on a Myr timescale.
Magnetic patches lose their angular momenta and move inward more quickly than their surroundings. In Section~\ref{sec:drift}, the inward drift speed will be compared with the theoretical prediction.

The behavior of the patches was consistent with that of the theoretical model by SU05. However, we note a difference from the theoretical model. In SU05, the magnetic fields of the mini-outflow roots are assumed to be strong and rigid against MRI, even around the disk midplane. However, in our model, the magnetic fields of the roots are subject to effective diffusivity in response to MRI turbulence. Therefore, the magnetic fields of the magnetic patches in both hemispheres move almost independently in our model. In other words, we demonstrated that the drifting motions of magnetic patches can occur asymmetrically in the Northern and Southern hemispheres.

We consider that the following processes are responsible for the formation of non-uniform $B_z$ around the disk surfaces. 
Poloidal magnetic fields are advected from the outside, while the poloidal fields accumulated around the center try to diffuse outward. The counter-transport of poloidal fields can create magnetic concentrations. We expect that this process will operate most efficiently near the center. Another important process is the fallback of outflows. As the magnetic pressure acceleration of outflows is gradual \citep{Kudoh_1998}, a portion of the outflows fall back onto the disk before their velocity exceeds the local escape velocity $v_{\rm esc}$ (Figures~\ref{fig:onset_bubble} and \ref{fig:example_magnetic_patch}), thereby complicating the magnetic structure. We notice that the fallback flows sometimes bring poloidal fields to the disk surfaces and induce magnetic concentrations. As shown in Figure~\ref{fig:example_magnetic_patch}, fallback flows produce a non-uniform poloidal field distribution (see the regions indicated by the thin arrows in Figure~\ref{fig:example_magnetic_patch}).

%%%%%%%%%%%%%%%
%
\section{Discussion}\label{sec:discussion}
%
%%%%%%%%%%%%%%%
\subsection{Global Story of the Spontaneous Formation of Magnetically driven Outflows}
\begin{figure}
\epsscale{1.0}
\plotone{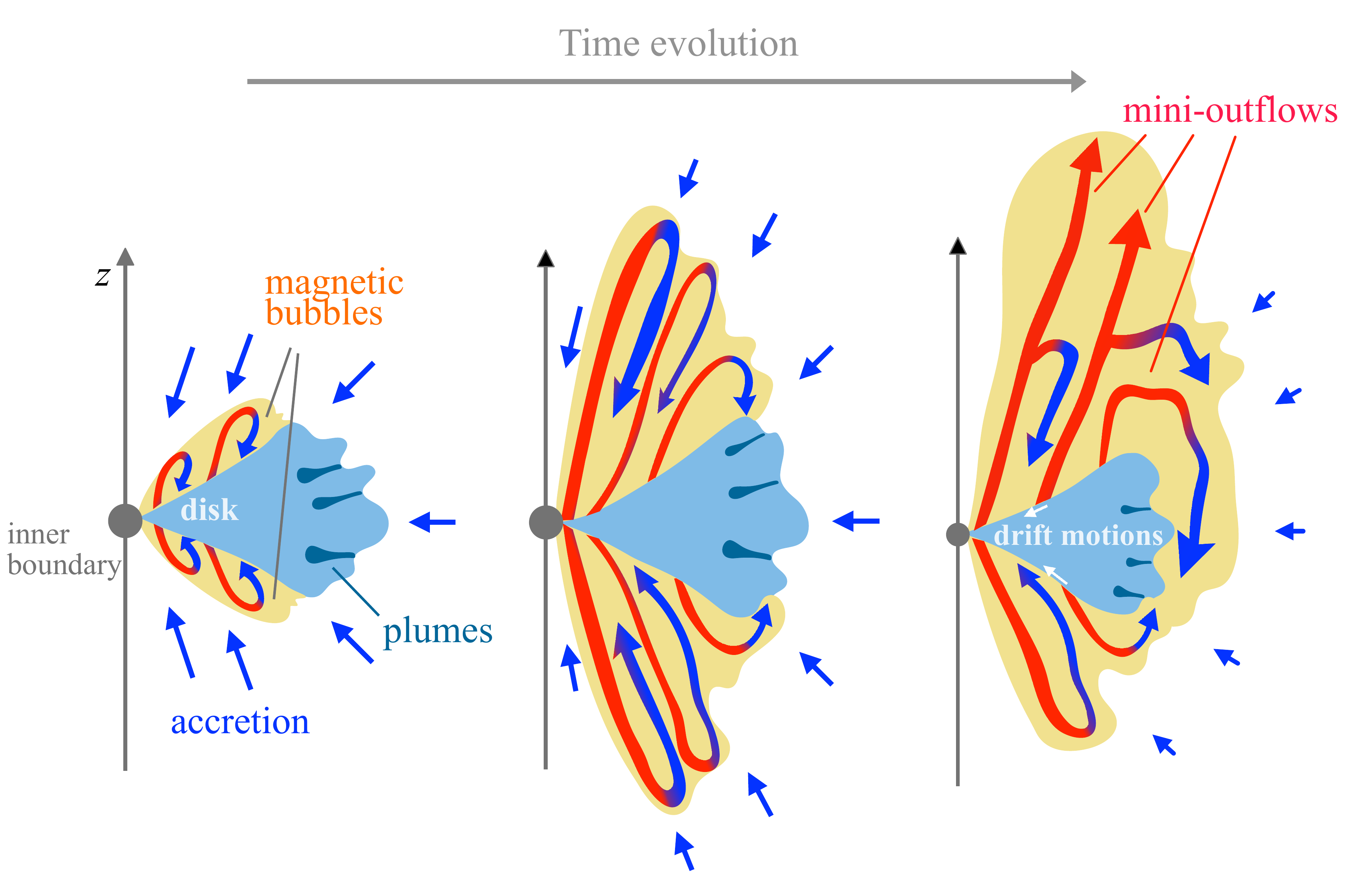}
\caption{Schematic of the development of the asymmetric outflow from the growing disk. 
\label{fig:schematic-diagram}}  %a-d:  Density_magpatch.py, e: Bpol_RSlice.py
\end{figure}

We briefly summarize the global story of the spontaneous formation of magnetically driven outflows. Figure~\ref{fig:schematic-diagram} shows a schematic of the development of the asymmetric outflow from the growing disk. The disk acquires poloidal magnetic fields from the accreting gas. The disk twists the poloidal fields to produce magnetic bubbles around the disk surfaces (left panel). At this stage, the bubbles are confined just around the disk, and outflows cannot emanate owing to the ram pressure of the accreting gas. 

At some point, the bubbles start to rapidly grow vertically against the accretion flow (middle panel of Figure~\ref{fig:schematic-diagram}). Inside the expanding bubbles, the gas does not suffer from the ram pressure of the accreting gas. As a result, the outflows can be extended vertically. The outflow structure is patchy because the distribution of the disk poloidal fields is uneven around the disk surfaces (Figure~\ref{fig:example_magnetic_patch}). The magnetic concentrations/patches drive mini-outflows. As mini-outflows are supersonic, they may produce internal shocks behind the bow shock. In magnetic bubbles, the toroidal component of the magnetic field is much larger than the poloidal component (Figure~\ref{fig:mag_va_beta}). Therefore, the outflows were driven mainly by the magnetic pressure gradient force. The simulation shows that the outflows help the magnetic bubbles expand by pushing the bubble surfaces outward. In Section~\ref{sec:growth_condition}, we show that rapid growth occurs when the disk radius exceeds a critical value. 

The north and south outflows are intrinsically asymmetric owing to the complex counter-flows excited by plumes around the outer edge of the disk (Figure~\ref{fig:vr_conv}). In this simulation, the north outflow started to grow slightly earlier than the south outflow. As a result, a portion of the accreting gas in the Northern hemisphere is directed toward the Southern hemisphere, leading to an increase in the mass accretion rate (Figure~\ref{fig:mass_inflow_rate}). The enhancement of the mass inflow rate decelerates the growth of the south outflow, enhancing the asymmetry between the two, as shown in the right panel of Figure~\ref{fig:schematic-diagram} (see also Figure~ \ref{fig:formation_outflow}).

Magnetically driven outflows carry angular momentum from disk surfaces. The bases of mini-outflows (magnetic patches) lose their angular momenta more rapidly than their surroundings. As a result, the bases show drift motions (Figures~\ref{fig:example_magnetic_patch} and \ref{fig:magnetic_patches_NS}), thereby introducing time variability in outflows. We examine the drift motions in more detail in Section~\ref{sec:drift}.

\subsection{Condition for the Growth of Magnetic Bubbles}\label{sec:growth_condition}
One of our goals is to understand how and when magnetically driven outflows are formed from the growing disk. We have seen that the development of magnetic bubbles precedes the outflow growth (Figure~\ref{fig:onset_bubble}). In this section, we discuss the growth condition of magnetic bubbles.

We expect that the magnetic bubbles will grow in size when the rate of the magnetic energy injection from the disk dominates the rate of kinetic energy injection by the accreting gas (hitting the bubbles). The magnetic energy injection results from the generation of $B_\phi$ from $B_z$ by the twisting motion of the disk. Considering this, we first estimate the (average) $B_z$ in the growing disk, $B_{\rm disk,qs}$. We then consider the field strength $B_z$ required for magnetic bubbles to grow in size, $B_{\rm disk,gr}$. 
The numerical results in Section~\ref{sec:results} imply that the condition for magnetic bubble growth,
i.e., $B_{\rm disk, qs} \approx B_{\rm disk, gr}$, 
is satisfied when the disk size exceeds
a critical radius (see Figures~\ref{fig:density_bline} and \ref{fig:onset_bubble}).
Here, we analytically estimate the critical radius as a function of the properties of the accretion flow. We first estimate $B_{\rm disk,qs}$ and then calculate $B_{\rm disk,gr}$.

We start from the induction equation for an axisymmetric disk in cylindrical coordinates $(R,\phi,z)$:
\begin{align}
    \frac{\partial B_z}{\partial t} = \frac{1}{R}\frac{\partial}{\partial R}\left[R(v_z B_R-v_R B_z)\right] - \frac{1}{R}\frac{\partial }{\partial R}\left[R\eta_{\rm eff}\left(\frac{\partial B_R}{\partial z}-\frac{\partial B_z}{\partial R}\right)\right].\label{eq:dBdt_1}
\end{align}
We integrate Equation~(\ref{eq:dBdt_1}) over the disk surface from the disk inner radius $R=r_{\rm in}$ to the disk radius $R=r_{\rm disk}$. 
\begin{align}
    \frac{\partial \Phi_{\rm disk}}{\partial t} \approx 2\pi r_{\rm disk} (v_z B_R - v_{R} B_{z})\rvert_{R=r_{\rm disk}} - 2 \pi r_{\rm disk} \eta_{\rm eff}(r_{\rm disk}) \left(\frac{\partial B_R}{\partial z}- \frac{\partial B_z}{\partial R} \right) \bigg\rvert_{R=r_{\rm disk}}, 
    \label{eq:dphidt_1}
\end{align}
where $\Phi_{\rm disk} \equiv \int_{r_{\rm in}}^{r_{\rm disk}}2\pi R B_z dR$ is the total flux of the poloidal magnetic fields in the disk (hereafter referred to as the total disk magnetic flux). We neglected the terms in the inner radius. Following previous studies \citep{Ogilvie2001ApJ,Okuzumi2014ApJ}, we divide Equation~(\ref{eq:dphidt_1}) by $\eta_{\rm eff}(r_{\rm disk})$ and integrate the equation with respect to $z$ within the range $-H^* \le z \le H^*$, where $H^*(>H)$ is the height just above the disk surface accretion. We define the following vertically averaged quantities:
\begin{align}
    \frac{1}{\eta_{\rm eff}^*(R)}&\equiv \frac{1}{2H^*}\int_{-H^*}^{H^*}\frac{dz}{\eta_{\rm eff}(R,z)}\\
    v_{R}^* &\equiv \frac{\eta_{\rm eff}^*(R)}{2H^*}\int_{-H^*}^{H^*}\frac{v_{\rm R}(R,z)}{\eta_{\rm eff}(R,z)}dz
\end{align}
and assume that $B_{z}(R,z)$ is nearly constant in the $z$ direction inside the disk.
The result of the integral is
\begin{align}
    \begin{split}
    \frac{1}{\eta_{\rm eff}^*}\frac{\partial \Phi_{\rm disk}}{\partial t}\approx 2\pi r_{\rm disk}\left( \frac{1}{2H^*}\int_{-H^*}^{H^*}\frac{v_z(R,z)B_R(R,z)}{\eta_{\rm eff}(R,z)}dz - \frac{v_{R}(R)^*}{\eta_{\rm eff}^*}B_z(R)\right)\bigg\rvert_{r=r_{\rm disk}}\\
    - 2\pi r_{\rm disk}\left( \frac{B_R(R,H^*)-B_R(R,-H^*)}{2H^*}-\frac{\partial B_z}{\partial R}\right)\bigg\rvert_{r=r_{\rm disk}}.\label{eq:dphidt_1a}
    \end{split}
\end{align}
Here, as the simulation shows that $|v_z|\ll |v_R|$ and $|B_R|\lesssim |B_z|$ for $|z|<H^*$, we neglect the first term on the right-hand side. In addition, $|B_R(R,H^*)-B_R(R,-H^*)|/2H^*\ll |\partial B_z/\partial R|$ around $R=r_{\rm disk}$. Therefore, we neglect the third term on the right-hand side. The third term determines the timescale of the relaxation of the poloidal field in the disk (see the timescale discussion below), but it is much smaller than the fourth term at the disk outer edge (note the rapid decrease in $B_z$ at the outer edge; Figure~\ref{fig:BzRslice}). As a result, the following equation was obtained:
\begin{align}
    \frac{\partial \Phi_{\rm disk}}{\partial t}\approx -2\pi r_{\rm disk}v_R^*(r_{\rm disk})B_{\rm acc} -2\pi r_{\rm disk}\eta_{\rm eff}^*(r_{\rm disk})\bigg\lvert \frac{\partial B_z}{\partial R}\bigg\rvert,\label{eq:dphidt_2}
\end{align}
where $B_{\rm acc}=B_z(r_{\rm disk})$ is the magnetic field strength outside the disk, which represents the field strength of the accreting gas.
The first term of Equation~(\ref{eq:dphidt_2}) on the right-hand side denotes the advection/injection of poloidal fields by the accreting gas ($ \dot{\Phi}_{\rm adv}$), and the second term indicates the diffusion of the disk poloidal fields from the disk body to the outside owing to the effective resistivity ($-\dot{\Phi}_{\rm diff}$).

We point out that the poloidal field is in a quasi-steady state on a disk viscous timescale and therefore $B_{\rm disk,qs}$ is characterized by the disk radius. When the disk radius exceeds the centrifugal radius, $r_{\rm disk}>R_{\rm c,\infty}$, an increase in the disk radius (or disk evolution) occurs on the viscous timescale $t_{\rm vis}(r_{\rm disk})=r_{\rm disk}^2/\nu(\rm r_{\rm disk})$. The relaxation timescale of the disk poloidal field is $t_{\rm diff}(r_{\rm disk})=r_{\rm disk}H(r_{\rm disk})/\eta_{\rm eff}^*(r_{\rm disk})$ \citep{lubow1994,lovelace2009}. This is the timescale for curved magnetic fields to become straight inside the disk, which is characterized by the third term in Equation~(\ref{eq:dphidt_1a}).
In our model, $t_{\rm diff}(r_{\rm disk})=(H(r_{\rm disk})/r_{\rm disk}) t_{\rm vis}(r_{\rm disk})$, where $H(r_{\rm disk})/r_{\rm disk}\approx 0.2$ for $r_{\rm disk}\approx 50 \,r_{\rm in}$ in our model. 
As the relaxation timescale of the poloidal fields $t_{\rm diff}$ is smaller than the disk evolution timescale $t_{\rm vis}$, we can assume that the disk poloidal fields are in a quasi-steady state on the timescale of $t_{\rm vis}(r_{\rm disk})$. In other words, the disk poloidal fields evolve on the timescale $t_{\rm vis}$ and are characterized by the disk radius $r_{\rm disk}$.
Figure~\ref{fig:BzRslice} shows that the distribution of the disk poloidal fields is steady after $t=17.8$~Myr. 
The relaxation timescale of the disk poloidal fields is estimated to be $t_{\rm diff}(r_{\rm disk}) \approx 10^5t_0\approx 16~{\rm Myr}$,
which is smaller than the time when the magnetic bubbles start to grow ($\sim 20$~Myr). Therefore, in the following, we consider the growth of magnetic bubbles under the assumption that $\partial \Phi_{\rm disk}/\partial t=0$. 

The poloidal field strength in the quasi-steady state 
$B_{\rm disk,s}$ is given by the balance between $\dot{\Phi}_{\rm adv}$ and $\dot{\Phi}_{\rm diff}$.
In our model, poloidal magnetic fields are brought to the disk by a nearly free-falling gas. Therefore, we approximate $v_{\rm R(r_{\rm disk})}\approx v_{\rm esc}(r_{\rm disk})$, where $v_{\rm esc}(r_{\rm disk})$ is the local escape velocity at a radius of $r_{\rm disk}$. We also consider $\partial B_z/\partial R\approx B_{\rm disk}/r_{\rm disk}$. Subsequently, we obtain
\begin{align}
    B_{\rm disk,qs}&\approx 1.3\times 10^{-1}B_{\rm 0} \left(\frac{\alpha}{10^{-2}}\right)^{-1} \left(\frac{c_s/v_{\rm K}(r_{\rm disk})}{0.2}\right)^{-2}\left(\frac{B_{\rm acc}}{B_{\rm B}}\right)\\
    &\propto r_{\rm disk}^{-1}.
\end{align}
Here, we assume that $\eta_{\rm eff}^*(r_{\rm disk})\approx \eta_{\rm eff}(r_{\rm disk},0)$ and the numerical result for $B_{\rm acc}$ (Figure~\ref{fig:BzRslice}). $B_{\rm acc}\approx B_{\rm B}\approx 5\times 10^{-5}B_{\rm 0}$ ($B_{\rm B}$ is the field strength at the Bondi radius) in our model. We have also used $H=\sqrt{2}c_{\rm s}/\Omega_{\rm K}.$ The estimated value is consistent with the numerical result shown in Figure~\ref{fig:BzRslice} within a factor of two.

Next, we estimate the field strength required for magnetic bubbles to grow in size, $B_{\rm disk,gr}$.
Just before the rapid growth, the magnetic bubbles have a nearly half-spherical shape with a radius of $\sim r_{\rm disk}$, and the plasma $\beta$ is smaller than unity (Figure~\ref{fig:onset_bubble}). For this reason, we assume that the surface area of a single bubble is $2\pi r_{\rm disk}^2$ and neglect the effect of the gas pressure. The rate of kinetic energy injection by the gas hitting the magnetic bubble can be expressed as 
\begin{align}
    \dot{E}_{\rm kin} &\approx 2\pi r_{\rm disk}^2 \frac{\rho_{\rm acc}}{2}v_{\rm esc}(r_{\rm disk})^3\\
    &= 2\sqrt{2}\pi \rho_{\rm acc} v_{\rm K}(r_{\rm disk})^3 r_{\rm disk}^2\label{eq:ekin},
\end{align}
where $\rho_{\rm acc}$ is the density of the accreting gas at $r=r_{\rm disk}$.
The rate of magnetic energy injection from the disk surface $\dot{E}_{\rm mag}$ can be written as
\begin{align}
    \dot{E}_{\rm mag}=\int_{r_{\rm in}}^{r_{\rm disk}}\frac{1}{t_{\rm tw}(R)}\frac{B_\phi(R,H^*)^2}{8\pi}2\pi w(R) R dR ,
\end{align}
where we assume that the magnetic energy is generated mainly within a disk surface layer with a thickness of $w(R)$ as a result of the amplification of the toroidal component $B_\phi$ on a timescale of $t_{\rm tw}(R)$. We consider that the generated magnetic energy is injected into the magnetic bubble from below.
Considering the results of previous 3D simulations \citep{Suzuki_Inutsuka_2014,Takasao_2018} and our simulation, the disk surface height was taken as $z=H^*(R)=2H(R)$, and the thickness was estimated to be $w(R)=2H(R)$.
The detailed description about $t_{\rm tw}(R)$ will be provided later.
When the magnetic bubbles start to expand, the magnetic energy density should be comparable to the gravitational energy density of the bubbles.
\begin{align}
    \frac{B_\phi(R,H^*)^2}{8\pi}\approx \frac{GM_{\rm BH}\rho_{\rm bub}}{\sqrt{R^2+(H^*)^2}}\approx\rho_{\rm bub}v_{\rm K}(R)^2,
\end{align}
where $\rho_{\rm bub}$ is the typical density of the magnetic bubbles, $v_{\rm K}(R)\equiv \sqrt{GM_{\rm BH}/R}$, and the approximation $(H^*/R)^2\ll 1$ is used. As the simulation suggests that the $R$-dependence of $\rho_{\rm bub}$ is insignificant, we assume that $\rho_{\rm bub}$ is a constant.
Using the above relation, we can rewrite $\dot{E}_{\rm mag}$ as 
\begin{align}
     \dot{E}_{\rm mag}\approx \int_{r_{\rm in}}^{r_{\rm disk}} \frac{1}{t_{\rm tw}(R)}\rho_{\rm bub}v_{\rm K}^2 2 \pi R w(R)dR. \label{eq:emag2}
\end{align}

$t_{\rm tw}(R)$ is estimated as follows. We consider the induction equation for $B_\phi$.
\begin{align}
    \frac{\partial B_\phi}{\partial t} \approx \frac{\partial v_\phi}{\partial R}B_R + \frac{\partial v_\phi}{\partial z}B_z, \label{eq:dBphidt2}
\end{align}
where we neglect the compression terms, as they are unimportant in and around the disk. When the magnetic bubbles start developing around the disk surfaces, strong toroidal fields prevent the formation of fast radial flows around the disk surfaces. Therefore, the second term, the generation of $B_\phi$ from $B_z$ by the disk twisting motion, is the dominant term. For a cold disk in a steady state, $v_{\phi}(R,z)=\sqrt{GM_{\rm BH}}R/(R^2+z^2)^{3/4}$. Thus, $\partial v_{\phi}/\partial z(R,H^*)\approx 3v_{\rm K}(R)H^*/2R^2$, where $(H^*/R)^2\ll 1$.
In Equation~\ref{eq:dBphidt2}, we assume that $\partial B_\phi/\partial t \approx B_\phi/t_{\rm tw}$.
As a result, we can estimate $t_{\rm tw}$ as
\begin{align}
    t_{\rm tw}^{-1} \approx \frac{3H^*v_{\rm K}(R)}{2R^2}\frac{B_{\rm disk,gr}}{B_\phi}.\label{eq:t_tw}
\end{align}

Now we are ready to estimate $B_{\rm disk,gr}$. From $\dot{E}_{\rm mag}=\dot{E}_{\rm kin}$ and Equations~(\ref{eq:ekin}), (\ref{eq:emag2}), and (\ref{eq:t_tw}), we obtain $B_{\rm disk,gr}$ as
\begin{align}
    B_{\rm disk,gr}\approx \frac{2\sqrt{\pi}}{3}\rho_{\rm acc}\rho_{\rm bub}^{-1/2}v_{\rm K}(r_{\rm disk})^3 c_s^{-2}\label{eq:bdisk_b1}
\end{align}
We can relate $\rho_{\rm acc}$ to the density at the Bondi radius $\rho_{\rm B}$ by assuming quasi-spherical accretion.
\begin{align}
    4\pi r_{\rm disk}^2\rho_{\rm acc}v_{\rm esc}(r_{\rm disk}) &\approx 4\pi r_{\rm B}^2 \rho_{\rm B} v_{\rm esc}(r_{\rm B}).
\end{align}
The validity of this assumption of spherical accretion was confirmed for the scale at $r=r_{\rm disk}>R_{c,\infty}$ in the simulation.
From this relation, Equation~(\ref{eq:bdisk_b1}) gives
\begin{align}
    B_{\rm disk,gr} &\approx \frac{2\sqrt{\pi}}{3}\rho_{\rm B}\rho_{\rm bub}^{-1/2}v_{\rm K}(r_{\rm disk})^3 c_s^{-2}\left(\frac{r_{\rm B}}{r_{\rm disk}}\right)^{3/2}\\
    &\propto  r_{\rm disk}^{-3}
\end{align}

By comparing the disk size dependence of $B_{\rm disk,qs}$ and $B_{\rm disk,gr}$, we can expect that a critical radius, where $B_{\rm disk,qs}=B_{\rm disk,gr}$ exists. When the disk size is smaller than the critical radius $r_{\rm disk,c}$, the saturated field strength $B_{\rm disk,qs}$ is insufficient for growing the magnetic bubbles. Once the disk size exceeds $r_{\rm disk,c}$, the magnetic bubbles start to grow in size. $r_{\rm disk,c}$ is calculated as follows:
\begin{align}
    r_{\rm disk,c} &= \frac{1}{\sqrt{3}}\alpha^{1/2} \left( \frac{v_{\rm K}(r_{\rm B})}{v_{\it A}'}\right)^{1/2}\left(\frac{\rho_{\rm B}}{\rho_{\rm bub}}\right)^{1/2}r_{\rm B},\label{eq:critical_radius}
\end{align}
where $v_{\it A}' = B_{\rm acc}/\sqrt{4\pi \rho_{\rm B}}\approx B_{\rm B}/\sqrt{4\pi \rho_{\rm B}}$. Our model assumes that $\alpha=10^{-2}$, $r_{\rm B}=10^3r_{\rm in}=100$~pc, and the simulation results show that $B_{\rm acc}\approx 5\times 10^{-5}B_0\approx 3.5\times 10^{-2}$~$\mu$G (Figure~\ref{fig:BzRslice}), $\rho_{\rm bub}\approx 10^4\rho_0=10^{-19}~{\rm g~cm^{-3}}$, and $\rho_{\rm B}=\rho_0=10^{-23}~{\rm g~cm^{-3}}$. We therefore obtain $r_{\rm disk,c}\approx 1.4\times 10^2 r_{\rm in}\approx 14$~pc. Although this value is approximately two to three times larger than the disk size at the time of the rapid growth of the magnetic bubbles ($r_{\rm disk}\approx 60r_{\rm in}$ and $t\approx $20~Myr), the above argument provides a theoretical explanation for the growth condition of magnetic bubbles and therefore the condition of the outflow formation.
The condition for the critical disk size is based on physical quantities at the Bondi radius and bubble density, although the bubble density remains to be determined. We infer that the bubble density will be determined by the property of the MRI-driven disk wind, as the disk wind of this type is expected to play an essential role in the mass loading to the upper atmospheres and global outflows of MRI-turbulent disks \citep{suzuki2009ApJ,bai2013ApJ}.

The magnetic bubbles grow explosively once the disk radius exceeds the critical radius because the timescale of the energy injection by the twisting motion of the disk $t_{\rm tw}$ is much smaller than $t_{\rm diff}$ and $t_{\rm vis}$. After the expansion of the magnetic bubbles, magnetically driven outflows can extend into the bubbles from the disk surfaces without suffering from the ram pressure of the accreting flows.

\subsection{Drift Speed of Mini-outflow Bases}\label{sec:drift}
Following SU05, we semi-analytically estimated the drift speed of mini-outflow bases (magnetic patches) and compared it with the numerical result. In the calculation, the differences in the assumptions of the theoretical model are noted. 

We consider the angular momentum loss of a magnetic patch based on the magnetic torque around the disk surface. Unlike SU05, the magnetic patch is localized around the disk surface (i.e., the magnetic field of the patch is dynamically disconnected from the disk surface of the opposite side). Here, we assume that the bottom and top heights of the root of the magnetic patch are $z_{\rm pb}$ and $z_{\rm pt}$, respectively. The top height corresponds to the disk surface. 

The angular momentum equation for the gas inside the patch is
\begin{align}
    \rho \frac{D}{Dt}(R v_{\phi})=\frac{\partial}{\partial z}\left( \frac{RB_\phi B_z}{4\pi}\right),
\end{align}
where $D/Dt$ is the Lagrangian time derivative. We integrated this equation in the $z$ direction inside the patch. Using the surface density of the patch $\Sigma_{\rm p}={\int_{z_{\rm pb}}^{z_{\rm pt}}}\,\rho \,dz$ and assuming that $v_\phi$ is constant inside the patch, we obtain
\begin{align}
    \Sigma_{\rm p}\frac{D}{Dt}(R v_{\phi})=\left[ \frac{RB_\phi B_z}{4\pi}\right]_{z_{\rm pb}}^{z_{\rm pt}}\approx \frac{RB_{\phi,{\rm pt}} B_{z,{\rm pt}}}{4\pi},\label{eq:patch-ang1}
\end{align}
where $B_{\phi,{\rm pt}}$ and $B_{ z,{\rm pt}}$ are the values of $B_{\rm phi}$ and $B_{\rm z}$ at $z=z_{\rm pt}$, respectively. As the simulation indicates that $|B_z(R,z_{\rm pb})|\ll |B_z(R,z_{\rm pt})|$ and $|B_\phi(R,z_{\rm pb})|\sim |B_\phi(R,z_{\rm pt})|$, we neglect the torque at the bottom of the patch. Using the surface area of patch $A_{\rm p}$, we define the mass of patch $M_{\rm p}=\Sigma_{\rm p}A_{\rm p}$. Assuming that $M_{\rm p}$ is constant during the drift motion, we obtain from Equation~(\ref{eq:patch-ang1})
\begin{align}
    \frac{D}{Dt}(M_{\rm p}Rv_\phi) \approx -\left\lvert \frac{A_{\rm p} R B_{\phi,{\rm pt}} B_{z,{\rm pt}}}{4\pi}\right\rvert.
\end{align}
$J_{\rm p}\equiv M_{\rm p}Rv_\phi$ is the total angular momentum of the base of the magnetic patch.

We can rewrite $DJ_{\rm p}/Dt$ as follows:
\begin{align}
\frac{DJ_{\rm p}}{Dt} = \frac{DR}{Dt}\frac{DJ_{\rm p}(R)}{DR}\approx -\frac{M_{\rm p}v_{\rm dr,p}v_{\rm K}(R)}{2},\label{eq:patch-ang2}
\end{align}
where $v_{\rm dr,p}=-DR/Dt$ is the drift speed of the magnetic patch. Here, we assume that $v_{\phi}(R,z)\approx v_{\rm K}(R,0)=v_{\rm K}(R)$.
By combining Equations~(\ref{eq:patch-ang1}) and (\ref{eq:patch-ang2}), we obtain
\begin{align}
    v_{\rm dr,p}&\approx \frac{R\lvert B_{\phi,{\rm pt}} B_{z,{\rm pt}}\rvert}{2\pi \Sigma_{\rm p}v_{\rm K}(R)}\\
    &\approx 2.5~{\rm pc~Myr^{-1}}~\left( \frac{H/R}{0.1}\right)^{-1}\left(\frac{\rho_{\rm pb}}{10^{5}\rho_{0}}\right)^{-1}\left(\frac{v_{\rm K}}{v_{\rm K}(10r_0)}\right)^{-1}\left(\frac{B_{z,{\rm pt}}}{0.5B_0}\right) \left(\frac{B_{\phi, {\rm pt}}}{20B_0}\right)\label{eq:drift-speed}
\end{align}
where $\rho_{\rm pb}=\rho(10r_0,z_{\rm pb})$. The values were obtained from the numerical simulation. The estimated drift speed agrees with the numerical result within a factor of four (Figure~\ref{fig:magnetic_patches_NS}).
Our results indicate that the theoretical model of SU05 provides a reasonable estimate of the drift speed with that accuracy. Equation~(\ref{eq:drift-speed}) shows that the drift speed explicitly depends on the radius ($v_{\rm dr,p}\propto R^{3/2}$) and is expected to decrease as the base moves toward the center if the other parameters are fixed. However, Figure~\ref{fig:magnetic_patches_NS} shows that most magnetic patches exhibit approximately constant speeds. We conjecture that an increase in the field strength suppresses the reduction in drift speed.

\subsection{Does Expansion of Magnetic Bubbles Continue?}\label{sec:expansion}
Our data analysis is limited to the data before numerical instabilities appear around the inner boundary ($t>26.3$~Myr). As the magnetic bubbles have not reached the Bondi radius by the time, whether the bubbles affect the Bondi radius scale remains unclear from our simulation. Here, we discuss the possibility of further expansion.

Whether bubbles continue to expand will be determined by the competition between magnetic pressure in the bubbles and ram pressure of the accreting gas. We evaluate them for expanding bubbles. As magnetic bubbles are collimated by their strong toroidal fields (Figures~\ref{fig:mag_va_beta} and \ref{fig:formation_outflow}), expansion is mainly in the $z$ direction. Therefore, the magnetic field of magnetic bubbles expands as a spring extends. Let us define the vertical length of the bubble in a hemisphere as $d$. Considering the magnetic flux conservation of the toroidal fields inside the bubble, we obtain $B_\phi r_{\rm disk}d={\rm const.}$
This gives a scaling of magnetic pressure based on the toroidal field as
\begin{align}
p_{{\rm mag},\phi}=\frac{B_\phi^2}{8\pi}\propto d^{-2}.\label{eq:pmag}
\end{align}
The ram pressure at a distance from the center $d$ is estimated as
\begin{align}
    p_{\rm ram}=\rho v_{r}^2\propto d^{-2.5}\label{eq:pram}
\end{align}
by assuming the spherical free-falling accretion. Therefore, once magnetic bubbles start to grow, the ram pressure cannot stop expansion of magnetic bubbles.

\begin{figure}
\epsscale{0.8}
\plotone{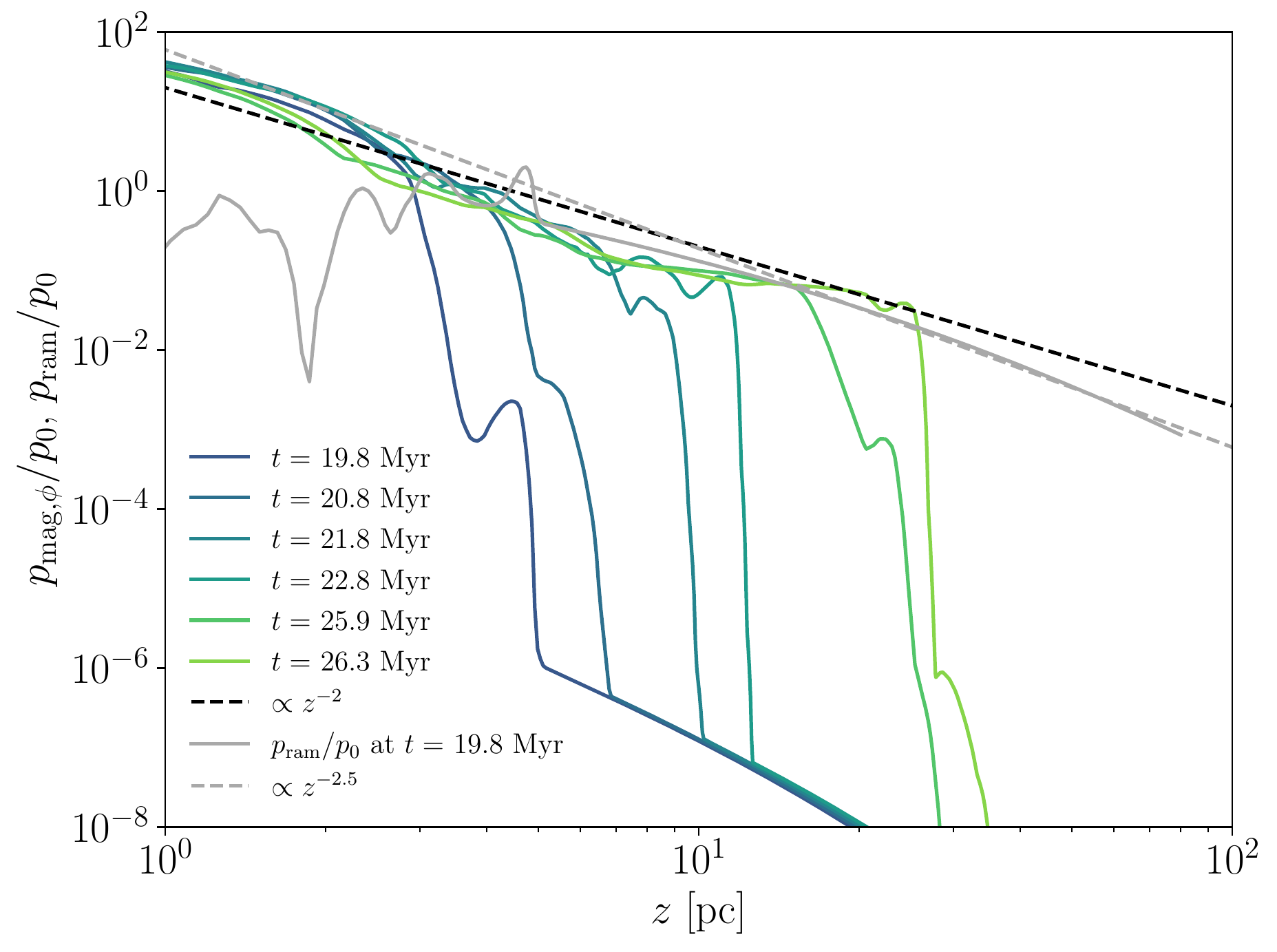}
\caption{Distributions of magnetic pressure based on the toroidal field $p_{{\rm mag},\phi}$ and ram pressure $p_{\rm ram}$ along the $z$ direction at $R=2$~pc. The values are normalized by $p_0 = \rho_0 v_0^2$. The colored lines denote $p_{{\rm mag},\phi}/p_0$ at different times, and the gray solid line indicates $p_{\rm ram}/p_0$ at $t=19.8$~Myr.
\label{fig:bubble_pmag}}
\end{figure}

The above scaling relations are confirmed by our simulation. Figure~\ref{fig:bubble_pmag} shows the distributions of the magnetic and ram pressures in the $z$ direction at $R=2$~pc. The values are normalized by $p_0 = \rho_0 v_0^2$. The solid gray line indicates the ram pressure $p_{\rm ram}$ when the magnetic bubbles start to grow, and the colored lines denote the magnetic pressure $p_{{\rm mag},\phi}$ at different times. The figure shows that both magnetic and ram pressures are consistent with the theoretical scaling relations (the black and gray dashed lines denote the theoretical scaling relations for $p_{{\rm mag},\phi}$ amd $p_{{\rm ram}}$, respectively). 

Our theoretical estimation suggests continuous growth of magnetic bubbles and potential feedback to a galactic scale. However, the difference between magnetic and ram pressures is insignificant for the scale range of our model. Therefore, in more realistic situations, fluctuations and anisotropy in the accreting gas may affect the evolution. The scaling relations (\ref{eq:pmag}) and (\ref{eq:pram}) indicate that the pressure difference will be more significant if there is a larger gap in scale between CND and the Bondi radius.

\subsection{Implications for the Molecular Outflows in NGC 1377}~\label{subsec:implications-molecular-outflows}
Using high-resolution ($2\times 3$~pc) ALMA 345~GHz observations of CO (3-2) and HCO$^+$ (4-3), \citet{aalto2020} observed a remarkable jet-like bipolar outflow of molecular gas in the radio-quiet, lenticular galaxy NGC 1377. The outflow is 150~pc long and its average diameter is 3--7~pc, indicating that the outflow is collimated. 
They also observed that the rotating molecular wind confined within the narrow angle range (50$^\circ$--70$^\circ$) surrounds the jet (see Figure~1 in their paper). 
The collimated bipolar outflow shows the line-of-sight velocity ``reversal'' along it. \citet{aalto2016} proposed that this structure may be a result of the precessing motion of the jet. The origin and driving mechanisms of these structures remain open questions. Considering the similarities to magnetically driven outflows in star-forming regions, \citet{aalto2020} speculated that the jet and winds are magneto-centrifugally driven, and they are powered by accretion in the galactic center. The dynamical mass inside the distance of 1.4~pc from the supermassive black hole is estimated to be $9\times 10^6 M_\odot$, equivalent to the mass of the supermassive black hole assumed herein. In this scenario, the jets and winds are considered to be driven by a large-scale poloidal magnetic field threading the (unresolved) nuclear disk.

Our numerical simulation demonstrated that gravitationally powered, magnetic-pressure-driven outflows can be generated as a natural consequence of weakly magnetized mass accretion toward a supermassive black hole. Although direct comparison between our numerical model and the observations toward NGC~1377 is not straightforward owing to the simplifications assumed herein, we provide some comments based on our numerical results. The bipolar outflow in the simulation was collimated by the hoop stress of the toroidal magnetic fields in the magnetic bubbles. The molecular jet in NGC~1377 could be collimated in a similar manner. The bipolar outflow in our simulation comprise multiple mini-outflows. Such mini-outflows are three-dimensionally seen as helical outflows if the bases of the outflows (magnetic patches) are localized in the azimuthal direction. The helical outflows can produce a line-of-sight velocity reversal. \citet{aalto2016} discussed possible origins of the precessing jet such as a warped disk \citep{greenhill2003}. We propose another scenario: helical outflows emanating from magnetic patches in nuclear disks. To test this scenario, more realistic 3D simulations were required.
Our current model simplifies radiative cooling and sets the minimum temperature of the gas to avoid numerical instabilities. The temperature floor is $4.5\times 10^4$~K, which is much higher than the dissociation temperature of the molecules. For this reason, our model cannot reproduce cold molecular outflows. Future models are required to address this issue.

From the discussion in Section \ref{sec:growth_condition}, we know that there is a critical radius of the nuclear disk where magnetic bubbles and outflows can grow. Equation~(\ref{eq:critical_radius}) states that the critical radius depends on the density and magnetic field strength at the Bondi radius and disk quantities (the viscous parameter $\alpha$ and the bubble density $\rho_{\rm bub}$). To test our theory, measuring the density and magnetic field strength at the Bondi radius scale and nuclear disk size is necessary. Our theoretical prediction is supported by observations if the nuclear disk size of NGC~1377 is observed to be larger than the critical radius for some fiducial values of $\alpha$ and $\rho_{\rm bub}$. The nuclear disk is currently not resolved with the beam size (i.e., $2\times 3$~pc) in the ALMA observations. Future observations with a higher spatial resolution will provide constraints on the nuclear disk size and reveal more detailed structures of outflows.

\subsection{Implications for the multi-phase outflows in nearby galaxies}
In this section, we briefly discuss the connection to observations. Molecular outflows in nearby star-forming galaxies and AGNs have been studied in detail \citep{fluetsch2019MNRAS.483.4586F,Runnoe2021-cx}. The driving mechanisms of molecular outflows have been under debate; they could be energy-driven \citep[e.g.,][]{king2010MNRAS.402.1516K}, momentum-driven \citep[e.g.,][]{king2015ARA&A..53..115K} or radiation pressure-driven \citep[e.g.,][]{wada2012radiation, Ishibashi2018-ow}. Most samples in \citet{fluetsch2019MNRAS.483.4586F} show that the ratio of the outflow momentum to the radiative momentum is larger than unity, suggesting the importance of central activities, such as AGNs or nuclear starbursts. Magnetic effects may be minor in those outflows.

Warm absorbers found in the soft X-ray spectra of AGNs \citep[e.g.,][]{Laha2014-nr, Laha2020-gs, blustin2005A&A...431..111B, Mizumoto2019-sh, ogawa2021} show similarities to the warm and diffuse outflows in our model. The typical velocity of observed warm absorbers ranges from 100 to 1000 km s$^{-1}$. Our results suggest that low-velocity ($\sim$100 km s$^{-1}$) species may be magnetically driven outflows emanating from parsec scale structures. \citet{blustin2005A&A...431..111B} indicated that warm absorbers are outflows emanating from the torus scale (a few to 5 pc) in 23 AGNs, which is consistent with our findings. \citet{Laha2020-gs} reported that 13 out of 20 sources do not show the variability in the column density N$_{\rm H}$ over years. Although MHD outflows in our model become non-steady because of complex outflows and fallback flows (see Figure~\ref{fig:example_magnetic_patch}), the variable timescale is $\sim 10^{5}$ yr. Therefore, we will not find time variability in observations of a single object, and the objects that show no variability is not inconsistent with the magnetically driven outflows. As the ionization degree and the temperature of outflows depend on the location of the outflow base and the cooling process, we will investigate the thermal structure with an updated model in future papers.

\subsection{Brief Comment on the Impact of the Parker Instability}
A brief comment on the impact of the Parker instability \citep{parker1966}, a 3D instability, will be provided. For the launch of escaping outflows, the magnetic energy density must be comparable to the gravitational energy density in the launching regions. However, \citet{Takasao_2018} showed using their 3D simulation that the Parker instability prevents an increase in the magnetic energy around the disk because the instability promotes the escape of magnetic fields. Here, we emphasize the difference in the accretion structure between the model of \citet{Takasao_2018} and our model. \citet{Takasao_2018} assumed a hydrostatic atmosphere outside the disk as an initial condition. However, in our model, the accretion flows coming from a large-scale hit the disk surfaces and bubbles. The free-falling accretion flows confined the magnetic fields around the disk surfaces until the magnetic energy density inside the bubbles was comparable to the gravitational energy density (Section~\ref{sec:growth_condition}). Therefore, confinement via free-falling accretion flows from a large solid angle eventually helps in the growth of magnetic-pressure-driven outflows. We expect that this is also true in three dimensions, where the Parker instability can occur because the confinement should be insensitive to the magnetic field geometry in the bubbles. Future detailed 3D modeling will reveal the impact of the Parker instability.

%%%%%%%%%%%%%%%
%
\section{Conclusions}\label{sec:conclusions}
%
%%%%%%%%%%%%%%%

We numerically and analytically investigated the spontaneous formation of magnetically driven outflows powered by rotating, weakly magnetized accretion flows in a galactic center. As this study requires a method that allows us to investigate the long-term evolution of the MRI-turbulent disk, we constructed an axisymmetric 2D disk model based on the effective viscosity and resistivity as a result of MRI turbulence (Section~2 and Appendix~\ref{appendix:MRI-model}). Our model demonstrates that the magnetically driven outflow extends inside the magnetic bubbles after the rapid expansion of the magnetic bubbles (Figures~\ref{fig:onset_bubble}, \ref{fig:formation_outflow}, and \ref{fig:schematic-diagram}). The disk acquires poloidal magnetic fields from the accreting gas. The disk poloidal field strength is determined by the balance between the injection and diffusion of the poloidal fields at the outer edge of the disk. This is our answer to key question~1: How does the central pc-scale disk acquire magnetic fields during its growth? Regarding key question~2: How and when are magnetically driven outflows powered by the growing disk? We derived the growth condition of magnetic bubbles for our model (Section~\ref{sec:growth_condition}), which corresponds to a necessary condition of magnetically driven outflow growth. Once the disk radius exceeds the critical radius, the outflow starts to rapidly grow. The outflow in our model was mainly driven by the magnetic pressure.

We studied the origin of the asymmetric outflow with mini-outflows (key question~3: What physical processes can produce substructures in outflows?). Our model demonstrates a bipolar outflow that is significantly asymmetric about the midplane (Figures~\ref{fig:formation_outflow} and \ref{fig:mass_inflow_rate}). The asymmetric structure originates from the complex flows excited by penetrating plumes around the outer edge of the growing disk (Figures~\ref{fig:vr_conv} and \ref{fig:density_bline}). The bipolar outflow comprises multiple mini-outflows (Figures~\ref{fig:onset_bubble} and \ref{fig:schematic-diagram}). The mini-outflows emanate from the magnetic concentrations (magnetic patches). Although we have not clearly identified the mechanism to produce the magnetic patches, we consider that both the radial and vertical transport of magnetic fields are responsible for the formation. The bases of mini-outflows show drift motions (Figures~\ref{fig:example_magnetic_patch} and \ref{fig:magnetic_patches_NS}). In Section~\ref{sec:drift}, we show that the drift speed can be estimated using a simple magnetic patch model based on the model established by \citet{Spruit_Uzdensky_2005}. The drift motions contribute to the time variability and inhomogeneity of the bipolar outflow.

It remains unclear if the outflows will become able to eventually escape from the galaxy. The speed of the majority of the outflow is smaller than the local escape velocity until the end of the simulation (Figure~\ref{fig:rotation_jet}). Also, our model ignore the gravitational potential of the bulge. In Section~\ref{sec:expansion}, using a theoretical argument, we suggested that magnetic bubbles will continue to grow and reach the Bondi radius. 
Indeed, at the end of the simulation, the magnetic bubbles are still expanding, and the mini-outflows inside them are growing in size. We have observed that the outflows help the magnetic bubbles expand by pushing the bubble surfaces outward. As the outflows do not escape, they eventually fail to outflow. However, successive outflows grow further because the size of the magnetic bubbles is larger than before. Therefore, our results indicate that magnetic bubbles and outflows co-evolve. We conjecture that this process will continue as long as the gas and magnetic fields are supplied via accretion. It will be interesting to see if outflows reach the Bondi radius and give feedback to galactic scale structures.

%\begin{acknowledgements}
We thank Kazumi Kashiyama, Takeru Suzuki and Kohei Inayoshi for their fruitful comments to the draft.
S.T. was supported by the JSPS KAKENHI grant Nos. JP18K13579 and JP21H04487.
K.W. was supported by JSPS KAKENHI Grant No. 21H04496.
Numerical computations were carried out on
Cray XC50 at the Center for Computational Astrophysics, National Astronomical Observatory of Japan. This work is partly based on Yuri Shuto's Master’s thesis (Kagoshima University, 2021).
%\end{acknowledgements}

\appendix
\section{Numerical Treatment for Inner Boundary}\label{appendix:inner-boundary}
Through many trials, we observed that numerical instabilities prevent long-term calculations as magnetic fields accumulate around the inner boundary. The numerical instabilities occur in the region where the plasma $\beta$ is much smaller than unity (i.e., strongly magnetized). To avoid numerical instabilities, we impose the following artificial resistivity just around the inner boundary ($r_{\rm in}< r \lesssim 3 r_{\rm in}$):
\begin{align}
    \eta_{\rm in}=\eta_{\rm in,0} \frac{1}{2}\left[-\tanh{\left( \frac{r-3r_{\rm in}}{0.1 r_{\rm in}} \right)+1}\right ].
\end{align}
The resistivity magnitude is determined in a manner such that the Lundquist number (or the magnetic Reynolds number), $S=Lv_{\it A}/\eta_{\rm in,0}$ is unity around the inner boundary and the artificial resistivity prevents the accumulation of fields there, where $L=r_{\rm in}$ is the typical length scale and $v_{\it A}=B/\sqrt{4\pi \rho}$ is the Alfv\'en speed. By trial and error, we adopt $\eta_{\rm in,0}=3\times 10^{-2}$ in the simulation unit. The inner resistive region is an order of magnitude smaller than the disk size, and $\sim 300$ times smaller than the Bondi radius. Therefore, we consider that the artificial resistivity will affect neither the magnetic fields for majority of the disk body nor the disk outflow.

\section{Radiative Cooling}\label{sec:cooling}
The accreting gas experiences an accretion shock when it approaches the centrifugal radius or collides with the rotationally supported disk gas. The accretion shock increases not only the density but also the specific entropy of the accreting gas $s$. We expect that the shocked gas will be cooled via radiative cooling and the CND will have a uniform temperature for simplicity. Therefore, we included it for the disk gas. Radiative cooling is not applied to the unshocked accreting gas; thus, we can focus on the magnetic effects of the disk outflow. The definition of the disk surface boundary is provided later. To model a cold, uniform-temperature CND, we performed simulations with the isothermal equation of state. However, the simulations were unsuccessful owing to numerical instabilities around accretion shocks and the inner boundary. This is the reason why we included the simplified radiative cooling for the disk gas.

We assume that the plasma is optically thin, and radiative cooling is mediated by collisional excitation of atoms and ions. Using the internal energy density $e_{\rm int}$, the radiative cooling timescale $t_{\rm rad}$ can be estimated by considering the change in the internal energy density caused by radiative cooling:
\begin{align}
    \frac{\partial e_{\rm int}}{\partial t}&=-n^2 \Lambda(T)\\
    t_{\rm rad}&=\frac{k_{\rm B}T}{\mu(\gamma-1)n \Lambda(T)}\\
    &\sim 9~{\rm yr}~\left(\frac{T}{10^{6}~{\rm K}}\right)\left(\frac{n}{10^5~{\rm cm^{-3}}}\right)^{-1}\left(\frac{\Lambda}{10^{-22}~{\rm erg~cm^{3}~s^{-1}}} \right)^{-1}.
\end{align}
where $n$ is the number density, $\mu$ is the mean molecular weight (assumed to be unity herein for simplicity). As the cooling timescale is much smaller than the timescale of the disk evolution, without any temperature floor, the temperature becomes much smaller than $10^{4}~{\rm K}$ quickly after the disk formation. However, simulating long-term ($>10$~Myr) evolution by spatially resolving the vertical structure of such a cold disk is numerically difficult. For this reason, we set the temperature floor as $T_{\rm floor}=10^{-3}T_0=4.5\times 10^{4}~{\rm K}$. The corresponding pressure scale height in the disk is $4.0r_{\rm in}$ at $r=R_{\rm c,\infty}=20r_{\rm in}$. In the numerical code, the temperature of the disk gas evolves in an operator-split manner by solving the following equation:
\begin{align}
    \frac{\partial T}{\partial t}=-\frac{T-T_{\rm floor}}{t_{\rm rad}}
\end{align}
Considering the functional form of the cooling function \citep{Sutherland_Dopita_1993,Wada_Papadopoulos_Spaans_2009}, we approximate $\Lambda(T)$ as 
\begin{align}
\Lambda(T)=\begin{cases}
10^{-22}~{\rm erg~cm^3~s^{-1}} & T \le 10^6~{\rm K}\\
10^{-23}~{\rm erg~cm^3~s^{-1}} & T > 10^6~{\rm K}.
\end{cases}
\end{align}
The temperature evolution is not sensitive to the choice of the functional form, as the CND temperature quickly becomes the temperature floor and uniform.

The accretion-shock-heated region is defined by density and specific entropy. We apply radiative cooling to the region where the density is larger than $10^3\rho_{\rm B}$, and the specific entropy is larger than $s/k_{\rm B}>-6.85$ in the simulation unit. This density is larger than that expected from the isotropic, free-fall gravitational contraction at $r=R_{\rm c,\infty}$ ($\sim 350\rho_{\rm B}$).

\section{Modeling Effective Resistivity and Viscosity in response to MRI turbulence}\label{appendix:MRI-model}

We emphasize the necessity of effective diffusivity in 2D axisymmetric models. After trials, we confirmed that in 2D axisymmetric, ideal MHD simulations, disk toroidal magnetic fields are subject to a continuous amplification and the disk plasma $\beta$ becomes close to or lower than unity as time proceeds, which is unlikely in reality. This failure is a direct consequence that 2D models cannot solve the magnetic reconnection of the toroidal component of magnetic fields, and the disk toroidal magnetic fields must be continuously amplified by the disk shearing motion.

Considering the above experience, we consider an approximate method to consider the dissipation of disk magnetic fields and the angular momentum transport via MRI turbulence. In this section, we describe our approach. The strength of the disk turbulence is often evaluated in terms of the disk viscosity parameter $\alpha$, which is approximately the ratio of the magnitude of the Maxwell tensor to the gas pressure. The effective viscosity $\nu_{\rm eff}$ is related to the viscosity parameter $\alpha$ as $\nu_{\rm eff} = \alpha c_{\rm s} H$.
As mentioned in Section~\ref{sec:effective_diff}, we expect $\alpha\sim 0.01$ for the MRI-saturated state.

MRI turbulence not only amplifies magnetic fields but also dissipates them via small-scale magnetic reconnection. Therefore, we expect an effective resistivity in response to the MRI turbulence. The effective resistivity $\eta_{\rm eff}$ was evaluated using 3D MHD simulations \citep{lesur2009}. The results demonstrate that 
\begin{align}
    \eta_{\rm eff}\sim \nu_{\rm eff}=\alpha c_{\rm s}H
\end{align}
Resistivity is intrinsically a tensor, and the values of the tensor components are generally different from each other. However, we deal with resistivity as a scalar here.

The effective viscosity and resistivity were assumed to operate only in the disk body. To switch on these diffusive terms only in the disk body, we define the following switching function:
\begin{align}
    f_{\rm disk}(\rho)=\frac{1}{2}\left[\tanh{\left( \frac{\rho -\rho_{\rm sw}}{\Delta \rho_{\rm sw}} \right)+1}\right ],
\end{align}
where $\rho_{\rm sw}=10^4\rho_{\rm B}$ and $\Delta \rho_{\rm sw}=10^3 \rho_{\rm B}$, respectively.
Subsequently, we write the effective resistivity and viscosity as follows:
\begin{align}
    \eta_{\rm eff}&=\nu_{\rm eff}=\alpha c_{\rm s}H f_{\rm disk}(\rho)
\end{align}
These diffusivities are turned off outside the disk. Therefore, the plasma outside the disk was tightly coupled with the magnetic field.

\begin{figure}
    \centering
    \includegraphics[width=18cm]{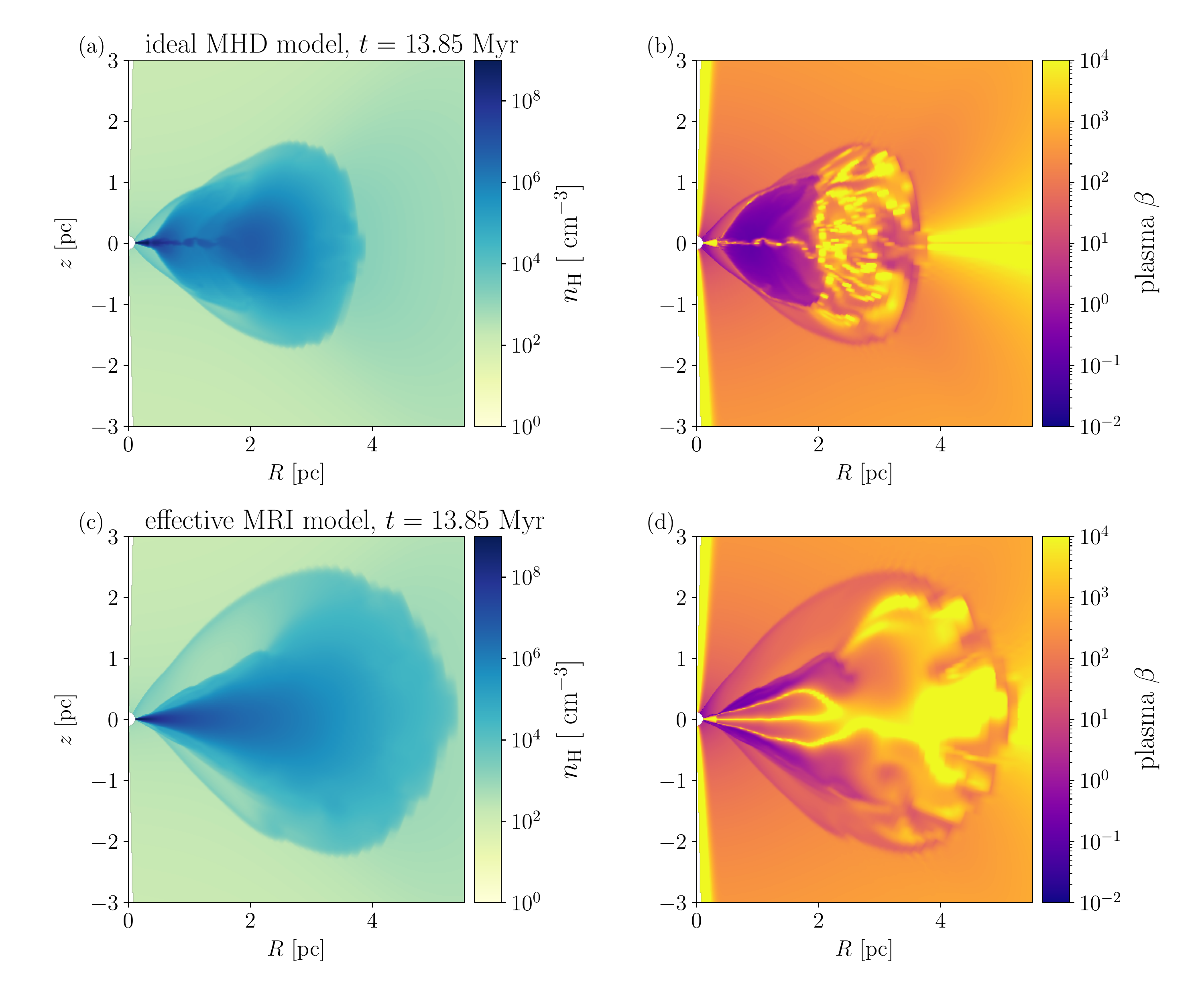}
 \caption
		{(top left) density distribution of a model without magnetic diffusion (top right) Same as left, but for plasma $\beta$ (bottom panels) Same as the top panels but for that with magnetic diffusion.}
      \label{fig:appendix}
\end{figure}

We demonstrate that our model of the effective resistivity and viscosity in response to MRI turbulence is useful for studying the long-term evolution of magnetized disks. Figure~\ref{fig:appendix} compares the disk structure of the ideal MHD model (left) and that of our model (right). Their initial and boundary conditions are the same as those in Section~\ref{sec:method}. In the ideal MHD model, a highly magnetized, low-$\beta$ disk is formed. The disk expands vertically owing to the magnetic pressure, and the gas around the midplane shows a clumpy distribution, which is not generally observed in 3D simulations with a similar setting. The accretion through the midplane is quenched at some radii owing to the strong magnetic fields. The formation of strongly magnetized disks is also observed in the axisymmetric, ideal MHD model of \citet{Romanova_2011}. On the contrary, the plasma $\beta$ in our model is maintained at $\sim$100 or larger, as observed in previous 3D simulations of MRI disks \citep{Suzuki_Inutsuka_2014,Takasao_2018}. The accretion in the disk was driven by the effective viscosity based on MRI turbulence. Therefore, our model does not suffer from the quenching of disk accretion by strong magnetic fields.

The reason why ideal MHD models produce disks with unrealistically strong magnetic fields is explained as follows. We indicate that the 2D axisymmetric ideal MHD models lack two key physical processes that suppress disk field amplification. When the MRI operates, $B_\phi$ is amplified via disk shear. In three dimensions, the amplification is suppressed by magnetic reconnection of the toroidal fields \citep{Sano_Inutsuka_2001}, but this does not occur in the axisymmetric model. Another important process is the Parker instability. The amplified toroidal fields are ejected from the disk on an orbital timescale via the Parker instability (a magnetic buoyancy instability) when the plasma $\beta$ is comparable to unity \citep{Takasao_2018}. As the Parker instability for the toroidal field is an undular-mode instability in the azimuthal direction, it cannot occur in the axisymmetric model. As the above two processes cannot occur in the axisymmetric models, $B_\phi$ is continuously amplified without realistic suppression. To avoid unrealistic field amplification, we need an effective diffusion based on the MRI turbulence property as the one used herein. We claim that the 2D axisymmetric simulations without any effective diffusion can provide unrealistic results owing to the above-mentioned reason.

\section{Rotational Velocity Structure around Magnetic Bubbles}\label{appendix:torque}
\begin{figure}
    \centering
    \includegraphics[width=18cm]{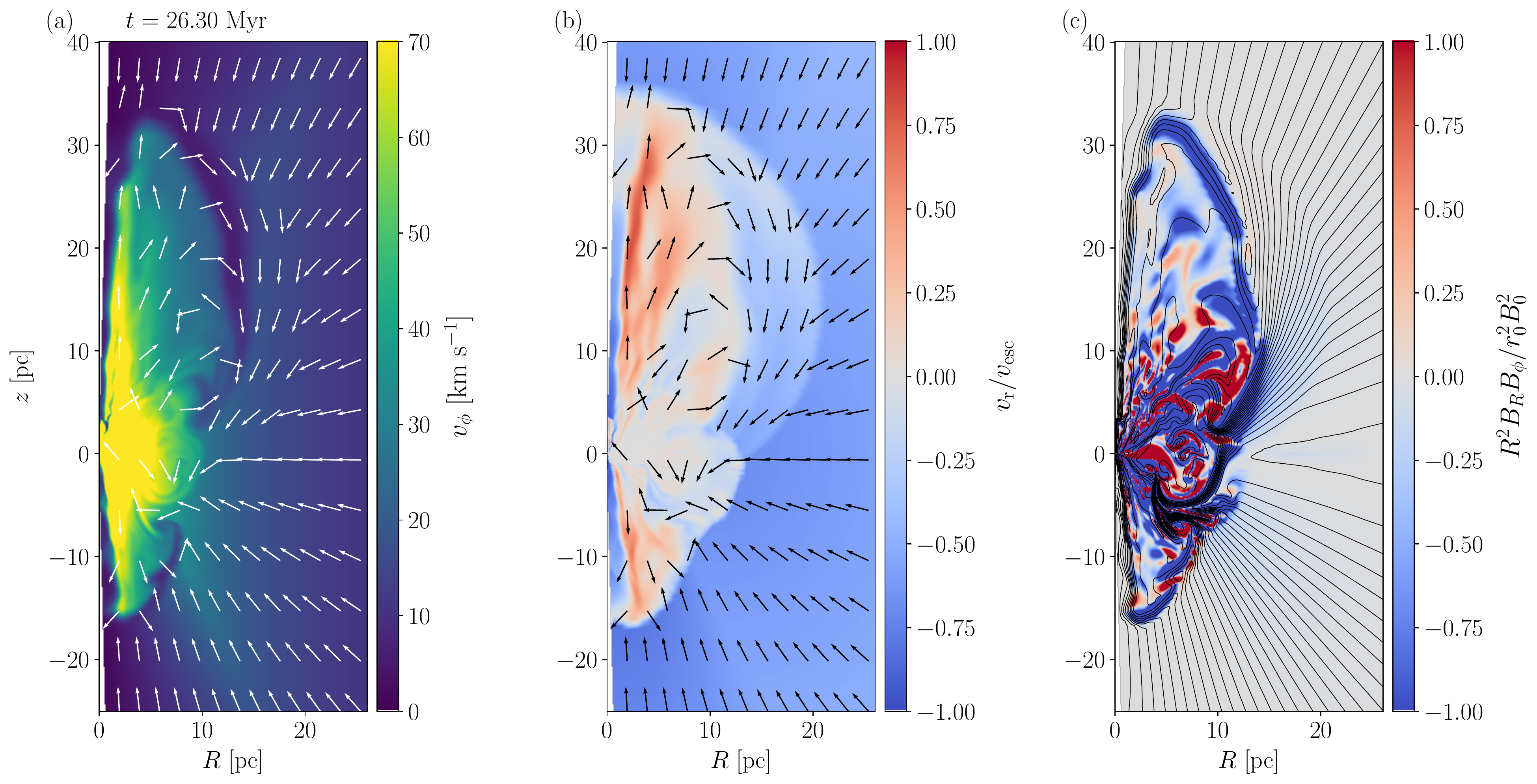}
    \caption{Rotational velocity structure around magnetic bubbles at the end of the simulation. Panels ~(a) and (b) show $v_\phi$ and $v_{r}/v_{\rm esc}$, respectively. $v_{\rm esc} = \sqrt{2G M_{BH}/r}$ is the local escape velocity. The arrows denote the direction of the velocity vectors (the arrow size does not indicate the speed). Panel~(c) shows $R^2 B_R B_\phi/r_0^2 B_0^2$ (see the text for more detail).}
      \label{fig:torque}
\end{figure}

Figure~\ref{fig:torque}(a) indicates a significant reduction in the rotational velocity at the outer edge of the bubbles. The reduction is caused by the magnetic force. We consider the equation of the angular momentum.
\begin{align}
    \frac{\partial}{\partial t}(\rho R v_\phi) = -\frac{1}{R}\frac{\partial}{\partial R}\left[R^2(\rho v_R v_\phi-B_R B_\phi)\right] - \frac{\partial }{\partial z}\left[R(\rho v_\phi v_z - B_\phi B_z)\right].\label{eq:torque}
\end{align}
As the magnetic torque is mainly caused by the change in the magnetic structure in the $R$ direction, $(1/R)\partial (R^2 B_R B_\phi)/\partial R$ is the dominant term on the right-hand side of Equation~(\ref{eq:torque}). Panel~(c) shows that $R^2 B_R B_\phi/r_0^2B_0^2$ takes a large negative value around the outer edge of the expanding bubbles, indicating that a large torque is produced there. The difference in the rotational velocity inside and outside the bubbles generates kinks in magnetic fields, thereby producing the magnetic torque. The accreting flows around the deceleration region receive angular momentum and slightly move outward (see the direction of velocity vectors indicated by arrows). However, they eventually accrete onto the disk (Panel~(b)).

\section{Energy Conversion Efficiency}\label{sec:energy_conversion}
The fraction of the accretion energy used to build up the magnetic bubbles and the formation of outflows is of interest. 
Therefore, we estimated the conversion efficiency of the accretion energy to the magnetic energy of the magnetic bubbles. First, we define the mean bubble density, $\bar{\rho}_{\rm bub}$
\begin{align}
    \bar{\rho}_{\rm bub}\equiv \frac{1}{{\rm V_{\rm bub}}}\int_{\rm V_{\rm bub}} \rho dV
\end{align}
where $\rm V_{\rm bub}$ is the volume of magnetic bubbles. The volume is defined as the region where $r \le r_{\rm disk}$ and $\rho \le \rho_{\rm bub}$.
$\bar{\rho}_{\rm bub}$ is typically much smaller than $\rho_{\rm bub}$, as the latter denotes the bubble density around the disk surfaces.
We also define
\begin{align}
    E_{\rm mag}&\equiv \int_{\rm V_{\rm bub}}\frac{B_\phi^2}{8\pi}dV \label{eq:emag_bubble}\\
    E_{\rm acc}&\equiv \frac{GM_{\rm BH}\dot{M}_{\rm B}}{r_{\rm disk}}t_{\rm vis}(r_{\rm disk}),\label{eq:eacc}
\end{align}
where $E_{\rm mag}$ represents the magnetic energy of the bubbles, and $E_{\rm acc}$ denotes the accretion energy released during the viscous timescale of the disk $t_{\rm vis}(r_{\rm disk})$. We take this timescale, considering that the disk radius increases on the viscous timescale. $\dot{M}_{\rm B}$ denotes the Bondi accretion rate. Using the above definitions, we aim to calculate $E_{\rm mag}/E_{\rm acc}$.
As the magnetic energy density is comparable to the gravitational energy density inside the bubbles, we rewrite Equation~\ref{eq:emag_bubble} as
\begin{align}
    E_{\rm mag}&\approx \int_{\rm V_{\rm bub}} \frac{GM_{\rm BH}\rho}{r} dV\\
    &\approx \frac{4\pi r_{\rm disk}^3}{3}\bar{\rho}_{\rm bub}\cdot\frac{GM_{\rm BH}}{r_{\rm disk}}.\label{eq:emag_bubble2}
\end{align}

Combining Equations~(\ref{eq:eacc}) and (\ref{eq:emag_bubble2}), we obtain
\begin{align}
    \frac{E_{\rm mag}}{E_{\rm acc}}&\approx \frac{M_{\rm bub}}{\dot{M}_{\rm B}t_{\rm vis}(r_{\rm disk})}\\
    &\approx 10^{-4}\left( \frac{M_{\rm bub}}{3\times 10^4M_{\odot}}\right) \left( \frac{\dot{M}_{\rm B}}{0.4~{M_{\odot}~{\rm yr}^{-1}}}\right)^{-1} \left( \frac{t_{\rm vis}}{1.6\times 10^{2}~{\rm Myr}}\right)^{-1},\label{eq:energy_ratio}
\end{align}
where $M_{\rm bub}=4\pi r_{\rm disk}^3\bar{\rho}_{\rm bub}/3$ is the total mass of the magnetic bubbles. The fiducial values are obtained from the numerical simulation approximately at the time of the rapid growth of the magnetic bubbles ($t\approx 20$~Myr) and $r_{\rm disk}=60\, r_{\rm in}=6$~pc. We also numerically calculated $E_{\rm mag}/E_{\rm acc}$ using Equations~(\ref{eq:emag_bubble}) and (\ref{eq:eacc}) and confirmed that the ratio is approximately $1.3\times 10^{-4}$, which is very close to the estimated value in Equation~(\ref{eq:energy_ratio}). The energy ratio can be written as the ratio of the bubble mass to the mass accreted at the Bondi accretion rate during the viscous timescale. The results show that approximately 0.01\% of the accretion energy was used for the formation of magnetic bubbles during the disk evolution timescale. As our estimate of the critical radius is consistent with the numerical result within a factor of a few, the accuracy of our theory is considered to be approximately 0.01\% in terms of the energetics. The dependence of the energy conversion efficiency on the physical condition at the Bondi radius can be determined by combining Equations~(\ref{eq:critical_radius}) and (\ref{eq:energy_ratio}), although the mechanisms that determine $\rho_{\rm bub}$ and the disk temperature (or the sound speed in the disk) will vary depending on the system of interest.

%\bibliography{MHDoutflow_AGN01,wada}{}
\bibliography{main_arxiv}{}

\begin{thebibliography}{}
\expandafter\ifx\csname natexlab\endcsname\relax\def\natexlab#1{#1}\fi
\providecommand{\url}[1]{\href{#1}{#1}}
\providecommand{\dodoi}[1]{doi:~\href{http://doi.org/#1}{\nolinkurl{#1}}}
\providecommand{\doeprint}[1]{\href{http://ascl.net/#1}{\nolinkurl{http://ascl.net/#1}}}
\providecommand{\doarXiv}[1]{\href{https://arxiv.org/abs/#1}{\nolinkurl{https://arxiv.org/abs/#1}}}

\bibitem[{{Aalto} {et~al.}(2016){Aalto}, {Costagliola}, {Muller}, {Sakamoto},
  {Gallagher}, {Dasyra}, {Wada}, {Combes}, {Garc{\'\i}a-Burillo}, {Kristensen},
  {Mart{\'\i}n}, {van der Werf}, {Evans}, \& {Kotilainen}}]{aalto2016}
{Aalto}, S., {Costagliola}, F., {Muller}, S., {et~al.} 2016, \aap, 590, A73,
  \dodoi{10.1051/0004-6361/201527664}

\bibitem[{{Aalto} {et~al.}(2017){Aalto}, {Muller}, {Costagliola}, {Sakamoto},
  {Gallagher}, {Falstad}, {K{\"o}nig}, {Dasyra}, {Wada}, {Combes},
  {Garc{\'\i}a-Burillo}, {Kristensen}, {Mart{\'\i}n}, {van der Werf}, {Evans},
  \& {Kotilainen}}]{aalto2017}
{Aalto}, S., {Muller}, S., {Costagliola}, F., {et~al.} 2017, \aap, 608, A22,
  \dodoi{10.1051/0004-6361/201730650}

\bibitem[{{Aalto} {et~al.}(2020){Aalto}, {Falstad}, {Muller}, {Wada},
  {Gallagher}, {K{\"o}nig}, {Sakamoto}, {Vlemmings}, {Ceccobello}, {Dasyra},
  {Combes}, {Garc{\'\i}a-Burillo}, {Oya}, {Mart{\'\i}n}, {van der Werf},
  {Evans}, \& {Kotilainen}}]{aalto2020}
{Aalto}, S., {Falstad}, N., {Muller}, S., {et~al.} 2020, \aap, 640, A104,
  \dodoi{10.1051/0004-6361/202038282}

\bibitem[{{Anderson} {et~al.}(2003){Anderson}, {Li}, {Krasnopolsky}, \&
  {Blandford}}]{anderson2003ApJ}
{Anderson}, J.~M., {Li}, Z.-Y., {Krasnopolsky}, R., \& {Blandford}, R.~D. 2003,
  \apjl, 590, L107, \dodoi{10.1086/376824}

\bibitem[{{Arce} {et~al.}(2007){Arce}, {Shepherd}, {Gueth}, {Lee}, {Bachiller},
  {Rosen}, \& {Beuther}}]{arce2007}
{Arce}, H.~G., {Shepherd}, D., {Gueth}, F., {et~al.} 2007, in Protostars and
  Planets V, ed. B.~{Reipurth}, D.~{Jewitt}, \& K.~{Keil}, 245.
\newblock \doarXiv{astro-ph/0603071}

\bibitem[{{Bai} \& {Stone}(2013)}]{bai2013ApJ}
{Bai}, X.-N., \& {Stone}, J.~M. 2013, \apj, 767, 30,
  \dodoi{10.1088/0004-637X/767/1/30}

\bibitem[{Balbus \& Hawley(1991)}]{Balbus_Hawley_1991}
Balbus, S.~A., \& Hawley, J.~F. 1991, The Astrophysical Journal, 376, 214,
  \dodoi{10.1086/170270}

\bibitem[{Beckwith {et~al.}(2009)Beckwith, Hawley, \& Krolik}]{Beckwith_2009}
Beckwith, K., Hawley, J.~F., \& Krolik, J.~H. 2009, The Astrophysical Journal,
  707, 428, \dodoi{10.1088/0004-637x/707/1/428}

\bibitem[{Bjerkeli {et~al.}(2016)Bjerkeli, van~der Wiel, Harsono, Ramsey, \&
  J{\o}rgensen}]{bjerkeli2016resolved}
Bjerkeli, P., van~der Wiel, M.~H., Harsono, D., Ramsey, J.~P., \& J{\o}rgensen,
  J.~K. 2016, Nature, 540, 406

\bibitem[{{Blandford} \& {Payne}(1982)}]{1982MNRAS.199..883B}
{Blandford}, R.~D., \& {Payne}, D.~G. 1982, \mnras, 199, 883,
  \dodoi{10.1093/mnras/199.4.883}

\bibitem[{{Blustin} {et~al.}(2005){Blustin}, {Page}, {Fuerst},
  {Branduardi-Raymont}, \& {Ashton}}]{blustin2005A&A...431..111B}
{Blustin}, A.~J., {Page}, M.~J., {Fuerst}, S.~V., {Branduardi-Raymont}, G., \&
  {Ashton}, C.~E. 2005, \aap, 431, 111, \dodoi{10.1051/0004-6361:20041775}

\bibitem[{{Chan} \& {Krolik}(2017)}]{chan_krolik2017ApJ...843...58C}
{Chan}, C.-H., \& {Krolik}, J.~H. 2017, \apj, 843, 58,
  \dodoi{10.3847/1538-4357/aa76e4}

\bibitem[{{Cicone} {et~al.}(2014){Cicone}, {Maiolino}, {Sturm},
  {Graci{\'a}-Carpio}, {Feruglio}, {Neri}, {Aalto}, {Davies}, {Fiore},
  {Fischer}, {Garc{\'\i}a-Burillo}, {Gonz{\'a}lez-Alfonso}, {Hailey-Dunsheath},
  {Piconcelli}, \& {Veilleux}}]{2014A&A...562A..21C}
{Cicone}, C., {Maiolino}, R., {Sturm}, E., {et~al.} 2014, \aap, 562, A21,
  \dodoi{10.1051/0004-6361/201322464}

\bibitem[{{Combes} {et~al.}(2019){Combes}, {Garc{\'\i}a-Burillo}, {Audibert},
  {Hunt}, {Eckart}, {Aalto}, {Casasola}, {Boone}, {Krips}, {Viti}, {Sakamoto},
  {Muller}, {Dasyra}, {van der Werf}, \& {Martin}}]{combes2019A&A...623A..79C}
{Combes}, F., {Garc{\'\i}a-Burillo}, S., {Audibert}, A., {et~al.} 2019, \aap,
  623, A79, \dodoi{10.1051/0004-6361/201834560}

\bibitem[{{Dihingia} {et~al.}(2021){Dihingia}, {Vaidya}, \&
  {Fendt}}]{dihingia2021}
{Dihingia}, I.~K., {Vaidya}, B., \& {Fendt}, C. 2021, \mnras, 505, 3596,
  \dodoi{10.1093/mnras/stab1512}

\bibitem[{{Dorodnitsyn} \& {Kallman}(2017)}]{dorodnitsyn2017ApJ...842...43D}
{Dorodnitsyn}, A., \& {Kallman}, T. 2017, \apj, 842, 43,
  \dodoi{10.3847/1538-4357/aa7264}

\bibitem[{{Fabian}(2012)}]{fabian2012}
{Fabian}, A.~C. 2012, \araa, 50, 455,
  \dodoi{10.1146/annurev-astro-081811-125521}

\bibitem[{Ferriere(2009)}]{ferriere2009interstellar}
Ferriere, K. 2009, Astronomy \& Astrophysics, 505, 1183

\bibitem[{{Fluetsch} {et~al.}(2019){Fluetsch}, {Maiolino}, {Carniani},
  {Marconi}, {Cicone}, {Bourne}, {Costa}, {Fabian}, {Ishibashi}, \&
  {Venturi}}]{fluetsch2019MNRAS.483.4586F}
{Fluetsch}, A., {Maiolino}, R., {Carniani}, S., {et~al.} 2019, \mnras, 483,
  4586, \dodoi{10.1093/mnras/sty3449}

\bibitem[{{Frank} {et~al.}(2014){Frank}, {Ray}, {Cabrit}, {Hartigan}, {Arce},
  {Bacciotti}, {Bally}, {Benisty}, {Eisl{\"o}ffel}, {G{\"u}del}, {Lebedev},
  {Nisini}, \& {Raga}}]{frank2014}
{Frank}, A., {Ray}, T.~P., {Cabrit}, S., {et~al.} 2014, in Protostars and
  Planets VI, ed. H.~{Beuther}, R.~S. {Klessen}, C.~P. {Dullemond}, \&
  T.~{Henning}, 451, \dodoi{10.2458/azu\_uapress\_9780816531240-ch020}

\bibitem[{{Fukui} {et~al.}(2006){Fukui}, {Yamamoto}, {Fujishita}, {Kudo},
  {Torii}, {Nozawa}, {Takahashi}, {Matsumoto}, {Machida}, {Kawamura},
  {Yonekura}, {Mizuno}, {Onishi}, \& {Mizuno}}]{fukui2006}
{Fukui}, Y., {Yamamoto}, H., {Fujishita}, M., {et~al.} 2006, Science, 314, 106,
  \dodoi{10.1126/science.1130425}

\bibitem[{{Garc{\'\i}a-Burillo} {et~al.}(2019){Garc{\'\i}a-Burillo}, {Combes},
  {Ramos Almeida}, {Usero}, {Alonso-Herrero}, {Hunt}, {Rouan}, {Aalto},
  {Querejeta}, {Viti}, {van der Werf}, {Vives-Arias}, {Fuente}, {Colina},
  {Mart{\'\i}n-Pintado}, {Henkel}, {Mart{\'\i}n}, {Krips}, {Gratadour}, {Neri},
  \& {Tacconi}}]{garcia-burillo2019A&A...632A..61G}
{Garc{\'\i}a-Burillo}, S., {Combes}, F., {Ramos Almeida}, C., {et~al.} 2019,
  \aap, 632, A61, \dodoi{10.1051/0004-6361/201936606}

\bibitem[{García-Burillo {et~al.}(2021)García-Burillo, Alonso-Herrero,
  Almeida, González-Martín, Combes, Usero, Hönig, Querejeta, Hicks, Hunt,
  Rosario, Davies, Boorman, Bunker, Burtscher, Colina, Díaz-Santos, Gandhi,
  García-Bernete, García-Lorenzo, Ichikawa, Imanishi, Izumi, Labiano,
  Levenson, López-Rodríguez, Packham, Pereira-Santaella, Ricci, Rigopoulou,
  Rouan, Shimizu, Stalevski, Wada, \& Williamson}]{garciaburillo2021galaxy}
García-Burillo, S., Alonso-Herrero, A., Almeida, C.~R., {et~al.} 2021, The
  Galaxy Activity, Torus and Outflow Survey (GATOS) I. ALMA images of dusty
  molecular tori in Seyfert galaxies.
\newblock \doarXiv{2104.10227}

\bibitem[{{Greenhill} {et~al.}(2003){Greenhill}, {Booth}, {Ellingsen},
  {Herrnstein}, {Jauncey}, {McCulloch}, {Moran}, {Norris}, {Reynolds}, \&
  {Tzioumis}}]{greenhill2003}
{Greenhill}, L.~J., {Booth}, R.~S., {Ellingsen}, S.~P., {et~al.} 2003, \apj,
  590, 162, \dodoi{10.1086/374862}

\bibitem[{{Hawley} {et~al.}(2015){Hawley}, {Fendt}, {Hardcastle}, {Nokhrina},
  \& {Tchekhovskoy}}]{hawley2015SSRv}
{Hawley}, J.~F., {Fendt}, C., {Hardcastle}, M., {Nokhrina}, E., \&
  {Tchekhovskoy}, A. 2015, \ssr, 191, 441, \dodoi{10.1007/s11214-015-0174-7}

\bibitem[{{Hawley} {et~al.}(2013){Hawley}, {Richers}, {Guan}, \&
  {Krolik}}]{hawley2013}
{Hawley}, J.~F., {Richers}, S.~A., {Guan}, X., \& {Krolik}, J.~H. 2013, \apj,
  772, 102, \dodoi{10.1088/0004-637X/772/2/102}

\bibitem[{{Heckman} \& {Best}(2014)}]{heckman2014}
{Heckman}, T.~M., \& {Best}, P.~N. 2014, \araa, 52, 589,
  \dodoi{10.1146/annurev-astro-081913-035722}

\bibitem[{{Hirota} {et~al.}(2017){Hirota}, {Machida}, {Matsushita}, {Motogi},
  {Matsumoto}, {Kim}, {Burns}, \& {Honma}}]{hirota2017NatAs}
{Hirota}, T., {Machida}, M.~N., {Matsushita}, Y., {et~al.} 2017, Nature
  Astronomy, 1, 0146, \dodoi{10.1038/s41550-017-0146}

\bibitem[{{Ho}(2008)}]{2008ARA&A..46..475H}
{Ho}, L.~C. 2008, \araa, 46, 475,
  \dodoi{10.1146/annurev.astro.45.051806.110546}

\bibitem[{{Hsieh} {et~al.}(2018){Hsieh}, {Koch}, {Kim}, {Ho}, {Tang}, \&
  {Wang}}]{hsieh2018}
{Hsieh}, P.-Y., {Koch}, P.~M., {Kim}, W.-T., {et~al.} 2018, \apj, 862, 150,
  \dodoi{10.3847/1538-4357/aacb27}

\bibitem[{{Imanishi} {et~al.}(2020){Imanishi}, {Nguyen}, {Wada}, {Hagiwara},
  {Iguchi}, {Izumi}, {Kawakatu}, {Nakanishi}, \&
  {Onishi}}]{imanishi2020ApJ...902...99I}
{Imanishi}, M., {Nguyen}, D.~D., {Wada}, K., {et~al.} 2020, \apj, 902, 99,
  \dodoi{10.3847/1538-4357/abaf50}

\bibitem[{{Inayoshi} {et~al.}(2019){Inayoshi}, {Ichikawa}, {Ostriker}, \&
  {Kuiper}}]{inayoshi2019MNRAS}
{Inayoshi}, K., {Ichikawa}, K., {Ostriker}, J.~P., \& {Kuiper}, R. 2019,
  \mnras, 486, 5377, \dodoi{10.1093/mnras/stz1189}

\bibitem[{{Inayoshi} {et~al.}(2020){Inayoshi}, {Visbal}, \&
  {Haiman}}]{inayoshi2020}
{Inayoshi}, K., {Visbal}, E., \& {Haiman}, Z. 2020, \araa, 58, 27,
  \dodoi{10.1146/annurev-astro-120419-014455}

\bibitem[{{Ishibashi} \& {Fabian}(2015)}]{ishibashi_fabian2015}
{Ishibashi}, W., \& {Fabian}, A.~C. 2015, \mnras, 451, 93,
  \dodoi{10.1093/mnras/stv944}

\bibitem[{Ishibashi {et~al.}(2018)Ishibashi, Fabian, \&
  Maiolino}]{Ishibashi2018-ow}
Ishibashi, W., Fabian, A.~C., \& Maiolino, R. 2018.
\newblock \doarXiv{1801.09700}

\bibitem[{{Joos} {et~al.}(2012){Joos}, {Hennebelle}, \& {Ciardi}}]{joos2012}
{Joos}, M., {Hennebelle}, P., \& {Ciardi}, A. 2012, \aap, 543, A128,
  \dodoi{10.1051/0004-6361/201118730}

\bibitem[{{Kakiuchi} {et~al.}(2018){Kakiuchi}, {Suzuki}, {Fukui}, {Torii},
  {Enokiya}, {Machida}, \& {Matsumoto}}]{kakiuchi2018}
{Kakiuchi}, K., {Suzuki}, T.~K., {Fukui}, Y., {et~al.} 2018, \mnras, 476, 5629,
  \dodoi{10.1093/mnras/sty629}

\bibitem[{{Kataoka} {et~al.}(2021){Kataoka}, {Yamamoto}, {Nakamura}, {Ito},
  {Sofue}, {Inoue}, {Nakamori}, \& {Totani}}]{kataoka2021}
{Kataoka}, J., {Yamamoto}, M., {Nakamura}, Y., {et~al.} 2021, \apj, 908, 14,
  \dodoi{10.3847/1538-4357/abdb31}

\bibitem[{{Kato} {et~al.}(2008){Kato}, {Fukue}, \&
  {Mineshige}}]{kato2008bhad.book.....K}
{Kato}, S., {Fukue}, J., \& {Mineshige}, S. 2008, {Black-Hole Accretion Disks
  --- Towards a New Paradigm ---}

\bibitem[{{Kato} {et~al.}(2004){Kato}, {Mineshige}, \&
  {Shibata}}]{y-kato2004ApJ}
{Kato}, Y., {Mineshige}, S., \& {Shibata}, K. 2004, \apj, 605, 307,
  \dodoi{10.1086/381234}

\bibitem[{{Kawakatu} \& {Wada}(2008)}]{kawakatu_wada2008ApJ...681...73K}
{Kawakatu}, N., \& {Wada}, K. 2008, \apj, 681, 73, \dodoi{10.1086/588574}

\bibitem[{{King} \& {Pounds}(2015{\natexlab{a}})}]{king2015}
{King}, A., \& {Pounds}, K. 2015{\natexlab{a}}, \araa, 53, 115,
  \dodoi{10.1146/annurev-astro-082214-122316}

\bibitem[{{King} \& {Pounds}(2015{\natexlab{b}})}]{king2015ARA&A..53..115K}
---. 2015{\natexlab{b}}, \araa, 53, 115,
  \dodoi{10.1146/annurev-astro-082214-122316}

\bibitem[{{King}(2010)}]{king2010MNRAS.402.1516K}
{King}, A.~R. 2010, \mnras, 402, 1516, \dodoi{10.1111/j.1365-2966.2009.16013.x}

\bibitem[{{Kudoh} {et~al.}(1998){Kudoh}, {Matsumoto}, \&
  {Shibata}}]{Kudoh_1998}
{Kudoh}, T., {Matsumoto}, R., \& {Shibata}, K. 1998, \apj, 508, 186,
  \dodoi{10.1086/306377}

\bibitem[{{Kudoh} {et~al.}(2020){Kudoh}, {Wada}, \&
  {Norman}}]{kudoh_wada2020ApJ...904....9K}
{Kudoh}, Y., {Wada}, K., \& {Norman}, C. 2020, \apj, 904, 9,
  \dodoi{10.3847/1538-4357/abba39}

\bibitem[{Laha {et~al.}(2014)Laha, Guainazzi, Dewangan, Chakravorty, \&
  Kembhavi}]{Laha2014-nr}
Laha, S., Guainazzi, M., Dewangan, G.~C., Chakravorty, S., \& Kembhavi, A.~K.
  2014, Mon. Not. R. Astron. Soc., 441, 2613

\bibitem[{Laha {et~al.}(2020)Laha, Markowitz, Krumpe, Nikutta, Rothschild, \&
  Saha}]{Laha2020-gs}
Laha, S., Markowitz, A.~G., Krumpe, M., {et~al.} 2020.
\newblock \doarXiv{2005.06079}

\bibitem[{{Lesur} \& {Longaretti}(2009)}]{lesur2009}
{Lesur}, G., \& {Longaretti}, P.~Y. 2009, \aap, 504, 309,
  \dodoi{10.1051/0004-6361/200912272}

\bibitem[{{Lovelace} {et~al.}(2009){Lovelace}, {Rothstein}, \&
  {Bisnovatyi-Kogan}}]{lovelace2009}
{Lovelace}, R.~V.~E., {Rothstein}, D.~M., \& {Bisnovatyi-Kogan}, G.~S. 2009,
  \apj, 701, 885, \dodoi{10.1088/0004-637X/701/2/885}

\bibitem[{{Lubow} {et~al.}(1994){Lubow}, {Papaloizou}, \&
  {Pringle}}]{lubow1994}
{Lubow}, S.~H., {Papaloizou}, J.~C.~B., \& {Pringle}, J.~E. 1994, \mnras, 267,
  235, \dodoi{10.1093/mnras/267.2.235}

\bibitem[{{Lynden-Bell}(1996)}]{lynden-bell1996MNRAS}
{Lynden-Bell}, D. 1996, \mnras, 279, 389, \dodoi{10.1093/mnras/279.2.389}

\bibitem[{{Machida} {et~al.}(2013){Machida}, {Nakamura}, {Kudoh}, {Akahori},
  {Sofue}, \& {Matsumoto}}]{machida2013ApJ...764...81M}
{Machida}, M., {Nakamura}, K.~E., {Kudoh}, T., {et~al.} 2013, \apj, 764, 81,
  \dodoi{10.1088/0004-637X/764/1/81}

\bibitem[{{Machida} {et~al.}(2009){Machida}, {Matsumoto}, {Nozawak},
  {Takahashi}, {Fukui}, {Kudo}, {Torii}, {Yamamoto}, {Fujishita}, \&
  {Tomisaki}}]{machida2009}
{Machida}, M., {Matsumoto}, R., {Nozawak}, S., {et~al.} 2009, \pasj, 61, 411,
  \dodoi{10.1093/pasj/61.3.411}

\bibitem[{{Machida} {et~al.}(2008){Machida}, {Inutsuka}, \&
  {Matsumoto}}]{machida2008ApJ}
{Machida}, M.~N., {Inutsuka}, S.-i., \& {Matsumoto}, T. 2008, \apj, 676, 1088,
  \dodoi{10.1086/528364}

\bibitem[{Matsumoto {et~al.}(1996)Matsumoto, Uchida, Hirose, Shibata, Hayashi,
  Ferrari, Bodo, \& Norman}]{Matsumoto_1996}
Matsumoto, R., Uchida, Y., Hirose, S., {et~al.} 1996, The Astrophysical
  Journal, 461, 115, \dodoi{10.1086/177041}

\bibitem[{Miyoshi \& Kusano(2005)}]{Miyoshi_Kusano_2005}
Miyoshi, T., \& Kusano, K. 2005, Journal of Computational Physics, 208, 315,
  \dodoi{10.1016/j.jcp.2005.02.017}

\bibitem[{Mizumoto {et~al.}(2019)Mizumoto, Done, Tomaru, \&
  Edwards}]{Mizumoto2019-sh}
Mizumoto, M., Done, C., Tomaru, R., \& Edwards, I. 2019.
\newblock \doarXiv{1907.01447}

\bibitem[{{Morris} {et~al.}(2006){Morris}, {Uchida}, \& {Do}}]{morris2006}
{Morris}, M., {Uchida}, K., \& {Do}, T. 2006, \nat, 440, 308,
  \dodoi{10.1038/nature04554}

\bibitem[{{Morris}(2015)}]{morris2015}
{Morris}, M.~R. 2015, {Manifestations of the Galactic Center Magnetic Field},
  391, \dodoi{10.1007/978-3-319-10614-4\_32}

\bibitem[{{Netzer}(2015)}]{2015ARA&A..53..365N}
{Netzer}, H. 2015, \araa, 53, 365, \dodoi{10.1146/annurev-astro-082214-122302}

\bibitem[{{Nomura} \& {Ohsuga}(2017)}]{nomura_ohsuga2017}
{Nomura}, M., \& {Ohsuga}, K. 2017, \mnras, 465, 2873,
  \dodoi{10.1093/mnras/stw2877}

\bibitem[{{Nomura} {et~al.}(2020){Nomura}, {Ohsuga}, \& {Done}}]{nomura2020}
{Nomura}, M., {Ohsuga}, K., \& {Done}, C. 2020, \mnras, 494, 3616,
  \dodoi{10.1093/mnras/staa948}

\bibitem[{Ogawa {et~al.}(2021)Ogawa, Ueda, Wada, \& Mizumoto}]{ogawa2021}
Ogawa, S., Ueda, Y., Wada, K., \& Mizumoto, M. 2021

\bibitem[{{Ogilvie} \& {Livio}(2001)}]{Ogilvie2001ApJ}
{Ogilvie}, G.~I., \& {Livio}, M. 2001, \apj, 553, 158, \dodoi{10.1086/320637}

\bibitem[{{Ohsuga} \& {Mineshige}(2011)}]{ohsuga2011}
{Ohsuga}, K., \& {Mineshige}, S. 2011, \apj, 736, 2,
  \dodoi{10.1088/0004-637X/736/1/2}

\bibitem[{{Ohsuga} \& {Mineshige}(2014)}]{ohsuga_mineshige2014SSRv..183..353O}
---. 2014, \ssr, 183, 353, \dodoi{10.1007/s11214-013-0017-3}

\bibitem[{{Okuzumi} {et~al.}(2014){Okuzumi}, {Takeuchi}, \&
  {Muto}}]{Okuzumi2014ApJ}
{Okuzumi}, S., {Takeuchi}, T., \& {Muto}, T. 2014, \apj, 785, 127,
  \dodoi{10.1088/0004-637X/785/2/127}

\bibitem[{{Parker}(1966)}]{parker1966}
{Parker}, E.~N. 1966, \apj, 145, 811, \dodoi{10.1086/148828}

\bibitem[{{Proga} {et~al.}(2000){Proga}, {Stone}, \& {Kallman}}]{proga2000}
{Proga}, D., {Stone}, J.~M., \& {Kallman}, T.~R. 2000, \apj, 543, 686,
  \dodoi{10.1086/317154}

\bibitem[{Romanova {et~al.}(2011)Romanova, Ustyugova, Koldoba, \&
  Lovelace}]{Romanova_2011}
Romanova, M.~M., Ustyugova, G.~V., Koldoba, A.~V., \& Lovelace, R. V.~E. 2011,
  Monthly Notices of the Royal Astronomical Society, 416, 416–438,
  \dodoi{10.1111/j.1365-2966.2011.19050.x}

\bibitem[{Runnoe {et~al.}(2021)Runnoe, G{\"u}ltekin, Rupke, \&
  L{\'o}pez-Sepulcre}]{Runnoe2021-cx}
Runnoe, J.~C., G{\"u}ltekin, K., Rupke, D., \& L{\'o}pez-Sepulcre, A. 2021.
\newblock \doarXiv{2105.13460}

\bibitem[{Sano \& Inutsuka(2001)}]{Sano_Inutsuka_2001}
Sano, T., \& Inutsuka, S.-i. 2001, The Astrophysical Journal, 561, L179–L182,
  \dodoi{10.1086/324763}

\bibitem[{{Shibata} \& {Uchida}(1985)}]{Shibata_1985}
{Shibata}, K., \& {Uchida}, Y. 1985, \pasj, 37, 31

\bibitem[{{Sofue}(1977)}]{sofue1977}
{Sofue}, Y. 1977, \aap, 60, 327

\bibitem[{{Sofue} {et~al.}(2005){Sofue}, {Kigure}, \& {Shibata}}]{sofue2005}
{Sofue}, Y., {Kigure}, H., \& {Shibata}, K. 2005, \pasj, 57, L39,
  \dodoi{10.1093/pasj/57.5.L39}

\bibitem[{Spruit \& Uzdensky(2005)}]{Spruit_Uzdensky_2005}
Spruit, H.~C., \& Uzdensky, D.~A. 2005, The Astrophysical Journal, 629, 960,
  \dodoi{10.1086/431454}

\bibitem[{Stone {et~al.}(2020)Stone, Tomida, White, \&
  Felker}]{Stone_Tomida_White_Felker_2020}
Stone, J.~M., Tomida, K., White, C.~J., \& Felker, K.~G. 2020, The
  Astrophysical Journal Supplement Series, 249, 4,
  \dodoi{10.3847/1538-4365/ab929b}

\bibitem[{Sugimura {et~al.}(2018)Sugimura, Hosokawa, Yajima, Inayoshi, \&
  Omukai}]{Sugimura_Hosokawa_Yajima_Inayoshi_Omukai_2018}
Sugimura, K., Hosokawa, T., Yajima, H., Inayoshi, K., \& Omukai, K. 2018,
  Monthly Notices of the Royal Astronomical Society, 478, 3961–3975,
  \dodoi{10.1093/mnras/sty1298}

\bibitem[{Sutherland \& Dopita(1993)}]{Sutherland_Dopita_1993}
Sutherland, R.~S., \& Dopita, M.~A. 1993, The Astrophysical Journal Supplement
  Series, 88, 253, \dodoi{10.1086/191823}

\bibitem[{{Suzuki} {et~al.}(2015){Suzuki}, {Fukui}, {Torii}, {Machida}, \&
  {Matsumoto}}]{suzuki2015}
{Suzuki}, T.~K., {Fukui}, Y., {Torii}, K., {Machida}, M., \& {Matsumoto}, R.
  2015, \mnras, 454, 3049, \dodoi{10.1093/mnras/stv2188}

\bibitem[{Suzuki \& Inutsuka(2014)}]{Suzuki_Inutsuka_2014}
Suzuki, T.~K., \& Inutsuka, S. 2014, The Astrophysical Journal, 784, 121,
  \dodoi{10.1088/0004-637x/784/2/121}

\bibitem[{{Suzuki} \& {Inutsuka}(2009)}]{suzuki2009ApJ}
{Suzuki}, T.~K., \& {Inutsuka}, S.-i. 2009, \apjl, 691, L49,
  \dodoi{10.1088/0004-637X/691/1/L49}

\bibitem[{{Takasao} {et~al.}(2021){Takasao}, {Aoyama}, \&
  {Ikoma}}]{takasao2021}
{Takasao}, S., {Aoyama}, Y., \& {Ikoma}, M. 2021, arXiv e-prints,
  arXiv:2106.16113.
\newblock \doarXiv{2106.16113}

\bibitem[{Takasao {et~al.}(2018)Takasao, Tomida, Iwasaki, \&
  Suzuki}]{Takasao_2018}
Takasao, S., Tomida, K., Iwasaki, K., \& Suzuki, T.~K. 2018, The Astrophysical
  Journal, 857, 4, \dodoi{10.3847/1538-4357/aab5b3}

\bibitem[{Takasao {et~al.}(2019)Takasao, Tomida, Iwasaki, \&
  Suzuki}]{Takasao_2019}
---. 2019, The Astrophysical Journal, 878, L10,
  \dodoi{10.3847/2041-8213/ab22bb}

\bibitem[{Tchekhovskoy(2015)}]{Tchekhovskoy2015}
Tchekhovskoy, A. 2015, Launching of Active Galactic Nuclei Jets, ed.
  I.~Contopoulos, D.~Gabuzda, \& N.~Kylafis (Cham: Springer International
  Publishing), 45--82, \dodoi{10.1007/978-3-319-10356-3_3}

\bibitem[{{The Event Horizon Telescope Collaboration} {et~al.}(2021){The Event
  Horizon Telescope Collaboration}, Akiyama, Algaba, Alberdi, Alef, Anantua,
  Asada, Azulay, Baczko, Ball, Balokovi{\'c}, Barrett, Benson, Bintley,
  Blackburn, Blundell, Boland, Bouman, Bower, Boyce, Bremer, Brinkerink,
  Brissenden, Britzen, Broderick, Broguiere, Bronzwaer, Byun, Carlstrom, Chael,
  Chan, Chatterjee, Chatterjee, Chen, Chen, Chesler, Cho, Christian, Conway,
  Cordes, Crawford, Crew, Cruz-Osorio, Cui, Davelaar, De~Laurentis, Deane,
  Dempsey, Desvignes, Dexter, Doeleman, Eatough, Falcke, Farah, Fish, Fomalont,
  Alyson~Ford, Fraga-Encinas, Freeman, Friberg, Fromm, Fuentes, Galison,
  Gammie, Garc{\'\i}a, Gentaz, Georgiev, Goddi, Gold, G{\'o}mez,
  G{\'o}mez-Ruiz, Gu, Gurwell, Hada, Haggard, Hecht, Hesper, Ho, Ho, Honma,
  Huang, Huang, Hughes, Ikeda, Inoue, Issaoun, James, Jannuzi, Janssen, Jeter,
  Jiang, Jimenez-Rosales, Johnson, Jorstad, Jung, Karami, Karuppusamy,
  Kawashima, Keating, Kettenis, Kim, Kim, Kim, Kim, Kino, Koay, Kofuji, Koch,
  Koyama, Kramer, Kramer, Krichbaum, Kuo, Lauer, Lee, Levis, Li, Li, Lindqvist,
  Lico, Lindahl, Liu, Liu, Liuzzo, Lo, Lobanov, Loinard, Lonsdale, Lu,
  MacDonald, Mao, Marchili, Markoff, Marrone, Marscher, Mart{\'\i}-Vidal,
  Matsushita, Matthews, Medeiros, Menten, Mizuno, Mizuno, Moran, Moriyama,
  Moscibrodzka, M{\"u}ller, Musoke, Mej{\'\i}as, Michalik, Nadolski, Nagai,
  Nagar, Nakamura, Narayan, Narayanan, Natarajan, Nathanail, Neilsen, Neri, Ni,
  Noutsos, Nowak, Okino, Olivares, Ortiz-Le{\'o}n, Oyama, {\"O}zel, Palumbo,
  Park, Patel, Pen, Pesce, Pi{\'e}tu, Plambeck, PopStefanija, Porth, P{\"o}tzl,
  Prather, Preciado-L{\'o}pez, Psaltis, Pu, Ramakrishnan, Rao, Rawlings,
  Raymond, Rezzolla, Ricarte, Ripperda, Roelofs, Rogers, Ros, Rose,
  Roshanineshat, Rottmann, Roy, Ruszczyk, Rygl, S{\'a}nchez,
  S{\'a}nchez-Arguelles, Sasada, Savolainen, Peter~Schloerb, Schuster, Shao,
  Shen, Small, Sohn, SooHoo, Sun, Tazaki, Tetarenko, Tiede, Tilanus, Titus,
  Toma, Torne, Trent, Traianou, Trippe, van Bemmel, van Langevelde, van Rossum,
  Wagner, Ward-Thompson, Wardle, Weintroub, Wex, Wharton, Wielgus, Wong, Wu,
  Yoon, Young, Young, Younsi, Yuan, Yuan, Anton~Zensus, Zhao, \&
  Zhao}]{The_Event_Horizon_Telescope_Collaboration2021-id}
{The Event Horizon Telescope Collaboration}, Akiyama, K., Algaba, J.~C.,
  {et~al.} 2021, ApJL, 910, L12

\bibitem[{{Tomida} {et~al.}(2013){Tomida}, {Tomisaka}, {Matsumoto}, {Hori},
  {Okuzumi}, {Machida}, \& {Saigo}}]{tomida2013ApJ}
{Tomida}, K., {Tomisaka}, K., {Matsumoto}, T., {et~al.} 2013, \apj, 763, 6,
  \dodoi{10.1088/0004-637X/763/1/6}

\bibitem[{{Tomisaka}(1998)}]{tomisaka1998ApJ...502L.163T}
{Tomisaka}, K. 1998, \apjl, 502, L163, \dodoi{10.1086/311504}

\bibitem[{{Tsukamoto} {et~al.}(2018){Tsukamoto}, {Okuzumi}, {Iwasaki},
  {Machida}, \& {Inutsuka}}]{tsukamoto2018ApJ}
{Tsukamoto}, Y., {Okuzumi}, S., {Iwasaki}, K., {Machida}, M.~N., \& {Inutsuka},
  S. 2018, \apj, 868, 22, \dodoi{10.3847/1538-4357/aae4dc}

\bibitem[{{Uchida} \& {Shibata}(1985)}]{1985PASJ...37..515U}
{Uchida}, Y., \& {Shibata}, K. 1985, \pasj, 37, 515

\bibitem[{Wada(2012)}]{wada2012radiation}
Wada, K. 2012, The Astrophysical Journal, 758, 66

\bibitem[{Wada(2015)}]{wada2015obscuring}
---. 2015, The Astrophysical Journal, 812, 82

\bibitem[{Wada {et~al.}(2009)Wada, Papadopoulos, \&
  Spaans}]{Wada_Papadopoulos_Spaans_2009}
Wada, K., Papadopoulos, P.~P., \& Spaans, M. 2009, The Astrophysical Journal,
  702, 63–74, \dodoi{10.1088/0004-637x/702/1/63}

\bibitem[{{Williamson} {et~al.}(2020){Williamson}, {H{\"o}nig}, \&
  {Venanzi}}]{willamson2020}
{Williamson}, D., {H{\"o}nig}, S., \& {Venanzi}, M. 2020, \apj, 897, 26,
  \dodoi{10.3847/1538-4357/ab989e}

\bibitem[{{Yuan} \& {Narayan}(2014)}]{yuan2014}
{Yuan}, F., \& {Narayan}, R. 2014, \araa, 52, 529,
  \dodoi{10.1146/annurev-astro-082812-141003}

\bibitem[{Zhu \& Stone(2018)}]{Zhu_Stone_2018}
Zhu, Z., \& Stone, J.~M. 2018, The Astrophysical Journal, 857, 34,
  \dodoi{10.3847/1538-4357/aaafc9}

\end{thebibliography}
\bibliographystyle{aasjournal}
\end{document}